\documentclass[11pt]{article}
\usepackage[left=2.5cm,top=2.5cm,right=2cm,bottom=2.5cm]{geometry}
\usepackage{amsmath, amssymb}

\usepackage{graphicx,epsfig,epsf,color}
\usepackage[utf8x]{inputenc}
\usepackage{slashed}
\usepackage{feynmf}
\usepackage{graphicx}
\usepackage{graphics}
\usepackage{epstopdf}
\usepackage{subcaption}
\usepackage{amsfonts}
\usepackage[colorlinks=true, urlcolor=blue, linkcolor=blue, citecolor=blue]{hyperref}

\usepackage{adjustbox}
\usepackage{cite}
\usepackage{pdflscape}
\usepackage{lipsum}
\usepackage{float}

\begin{document}
 \date{}

\title{Bayesian Constraints on Inverse-Tangent Inflation with Constant-EOS Reheating and a Dynamical Reheating Analysis}
 \maketitle
 \begin{center}
\author{Mayur Abhisheki} \footnote{Corresponding author: p20240040@goa.bits-pilani.ac.in}
~Prasanta Kumar Das \footnote{pdas@goa.bits-pilani.ac.in} \\
 \end{center}

 \begin{center}
Department of Physics, Birla Institute of Technology and Science, \\
K. K. Birla Goa Campus, NH-17B, Zuarinagar, Sancoale, Goa 403726, India
 \end{center}
 \vspace*{0.25in}

\begin{abstract}
We perform a Bayesian inference analysis of an inflationary model based on an inverse-tangent potential, incorporating reheating dynamics in both constant and dynamical equation-of-state (DEOS) frameworks. Using \emph{Planck} and \emph{ACT} constraints on the scalar spectral index, we find preferred values $\kappa\simeq0.5-0.6$ and $N_k\simeq40-60$, leading to reheating temperatures $T_{RH}\sim10^{10}-10^{14}$ GeV and reheating durations $N_{RH}\sim3-36$ e-folds. Reheating weighted $H_0$ posteriors shift the \emph{Planck} inference towards the \emph{ACT} preferred region through the intrinsic $n_s-H_0$ degeneracy of the CMB likelihood.
In the DEOS framework, reheating with a constant decay rate yields $N_{RH}\simeq4-8$ e-folds and $T_{RH}\simeq10^{13}$ GeV, while a dynamical decay rate produces a strong dependence on the Yukawa coupling $y$, with $N_{RH}$  varying from $\mathcal{O}(30)$ to $\mathcal{O}(1)$ e-folds and the reheating temperature spanning $\sim10^{-2}-10^{14}$ GeV. Imposing inflation-reheating consistency significantly restricts the viable parameter space to a narrow region around $n_s\simeq0.9720-0.9725$. and $r\simeq0.026-0.060$, demonstrating that reheating dynamics provide a nontrivial bridge between early-universe inflation and late-time cosmological parameter inference.
\end{abstract}

{\bf Keywords:} Inflation, slow-roll parameters, spectral index parameters, e-fold, reheating. 

\section{Introduction}

The very early universe is believed to have undergone a phase of accelerated expansion known as inflation, which successfully addresses the horizon and flatness problems \cite{Guth,Linde} and provides the origin of primordial density perturbations that seeded cosmic structure formation \cite{Hawking,Guth2,Starobinsky2,Bardeen}. Precision measurements of the cosmic microwave background (CMB) by the \emph{Planck} \cite{Planck} and \emph{ACT} \cite{ACT, ACT2} collaborations have placed stringent constraints on inflationary observables such as the scalar spectral index $n_s$ and the tensor-to-scalar ratio $r$. These measurements allow inflationary models to be tested beyond background dynamics through comparison with observational data.

A subtle but crucial aspect of this comparison is the role of reheating. After inflation ends, the universe must transition to a radiation-dominated phase in order to reproduce the successful predictions of Big Bang Nucleosynthesis (BBN) \cite{KolbTurner,Weinberg}. This transition occurs through reheating, during which the oscillating inflaton field decays into lighter degrees of freedom \cite{Abbott, Dolgov, Albrecht2, Kofman, Traschen, Kofman2}. Importantly, the duration and dynamics of reheating determine the number of e-folds between horizon exit and the end of inflation, $N_k$. Since the inflationary predictions for $n_s$ and $r$ depend explicitly on $N_k$, the comparison between inflationary models and CMB data cannot be performed without specifying how reheating occurs.

In most studies, this degeneracy between inflationary parameters and reheating dynamics is treated heuristically by assuming a constant reheating equation of state $\omega_{RH}$ \cite{Dai, Munoz, Cook, Ueno}. However, such an assumption may not capture the true physics of reheating, where the relative contributions of the inflaton condensate and the produced radiation evolve continuously. A more realistic description requires incorporating a dynamical equation of state during reheating \cite{Ghosh}.

A key point is that CMB data does not directly constrain reheating parameters, but only the spectral index $n_s$. Since reheating determines $N_k$, which in turn determines $n_s$, observational constraints on $n_s$ can be propagated backward to place constraints on reheating physics. Conversely, any restriction on reheating dynamics selects specific regions within the CMB posterior. This provides a natural Bayesian bridge between early-universe reheating physics and late-time cosmological parameter inference.

In our previous work, we introduced an inflationary model based on an arctangent potential \cite{Mayur}
\[
V(\phi)=V_0\,\arctan^2(\kappa\phi/m_p),
\]
and demonstrated that its shape interpolates between plateau-like behavior at large field values and quadratic behavior near the origin. While the model was shown to be compatible with observational bounds in the $n_s$–$r$ plane, the comparison was performed by directly overlaying theoretical trajectories onto observational contours, without performing a statistical inference of the model parameters.

The purpose of the present work is to embed this inflationary model into a fully Bayesian framework that connects inflationary dynamics, reheating physics, and CMB parameter inference. We first construct an interpolated mapping from $(\kappa, N_k)$ to $(n_s, r)$ using numerically precomputed solutions and perform a Bayesian inference using \emph{Planck} constraints on $n_s$ and the upper bound on $r$ to obtain posterior distributions for the inflationary parameters themselves.

We then incorporate reheating physics in two complementary ways. First, we consider the conventional treatment with a constant reheating equation of state $\omega$, leading to posterior distributions in the extended parameter space $(\kappa, N_k, T_{RH}, N_{RH})$ entirely induced by \emph{Planck} constraints on $n_s$. We demonstrate that reheating parameters, although not directly observable, become probabilistically constrained through this mapping.

Next, we interface the inflationary–reheating posterior with the \emph{Planck} and \emph{ACT} MCMC chains for $\Lambda$CDM parameters. By mapping our posterior samples onto the CMB chains, we investigate how preferred reheating histories correlate with the inferred value of the Hubble constant $H_0$. This provides a direct and model-independent link between early-universe reheating dynamics and late-time cosmological parameter inference.

Finally, we extend the analysis to include a dynamical equation of state (DEOS) during reheating, obtained by solving the coupled evolution equations for the inflaton and radiation energy densities with a time-dependent decay rate $\Gamma(t)$. This allows the reheating temperature and duration to be computed directly from the field dynamics, leading to a more realistic and self-consistent description of perturbative reheating.

\section{Constant EOS Reheating Model}

For a given inflationary potential, the observables $n_s$ and $r$ are determined by the slow-roll parameters, \cite{Cline}
\begin{equation}
n_s \simeq 1 - 6\epsilon + 2\eta, \qquad
r \simeq 16\epsilon,
\end{equation}
where
\begin{equation}
\epsilon=\frac{m_p^2}{2}\left ( \frac{V'}{V} \right )^2, \quad 
\eta=m_p^2 \left ( \frac{V''}{V}\right).
\label{sr}
\end{equation}

These quantities depend on the value of the field at horizon exit, which in turn depends on the number of e-folds $N_k$. However, $N_k$ is not a free parameter: it depends on the post-inflationary expansion history and hence on reheating dynamics. Therefore, specifying the reheating behavior determines which values of $N_k$ and, consequently, which values of $n_s$, are allowed.

Since \emph{Planck} tightly constrains $n_s$, this automatically induces constraints on reheating parameters. In this sense, reheating quantities are not independently fitted, but their posterior distributions arise entirely from mapping the \emph{Planck} posterior of $n_s$ through the reheating relations.

Assuming that reheating can be described by an effective equation of state
\begin{equation}
p=\omega_{RH}\,\rho,
\end{equation}
the number of reheating e-folds is given by
\begin{equation}
N_{RH}=\ln\left(\frac{a_{RH}}{a_{end}}\right)
=\frac{1}{3(1+\omega_{RH})}\ln \left(\frac{\rho_{end}}{\rho_{RH}} \right).
\label{Nre}
\end{equation}

Using entropy conservation between reheating and today \cite{Cook}, one obtains
\begin{equation}
T_{RH}=\left ( \frac{43}{11g_{re}}  \right )^{1/3} 
\left(\frac{a_0T_0}{k}\right)H_ke^{-N_k}e^{-N_{RH}},
\label{Tre}
\end{equation}
and
\begin{equation}
N_{RH}=\frac{4}{1-3\omega_{RH}}\left[61.6-\ln\left(\frac{V_{end}^{1/4}}{H_k}\right)-N_k\right].
\label{Nre2}
\end{equation}

Equations \eqref{Tre} and \eqref{Nre2} explicitly show that reheating parameters $(T_{RH}, N_{RH})$ are functions of $N_k$. Since $N_k$ determines $n_s$, and $n_s$ is tightly constrained by \emph{Planck}, the \emph{Planck} posterior for $n_s$ induces a non-trivial posterior distribution in the reheating parameter space $\{\kappa, N_k, T_{RH}, N_{RH}\}$.


\section{Dynamical EOS Reheating Model} \label{Our Model}

The constant equation-of-state treatment of reheating, discussed in the previous section, provides a useful but idealized description of the post-inflationary dynamics. In realistic perturbative reheating scenarios, the inflaton oscillates about the minimum of its potential and gradually decays into lighter degrees of freedom. As a result, the relative contributions of the inflaton condensate and the radiation produced evolved continuously, leading to a time-dependent equation of state during reheating.

Once slow-roll conditions break down and the field approaches the minimum of the potential, the Klein–Gordon equation becomes
\begin{equation}
    \ddot{\phi} + 3H \dot{\phi} + V'(\phi) = 0.
\end{equation}
The field then undergoes coherent oscillations about the minimum of $V(\phi)$. To phenomenologically describe the decay of the inflaton into radiation, we introduce an additional friction term \cite{Peskin} $\Gamma \dot{\phi}$, giving
\begin{equation}
    \ddot{\phi} + 3H \dot{\phi} + \Gamma \dot{\phi} + V'(\phi) = 0.
    \label{reos2}
\end{equation}

The inflaton energy density and pressure are
\begin{equation}
\rho_\phi = \frac{\dot{\phi}^2}{2} + V(\phi), \qquad
p_\phi = \frac{\dot{\phi}^2}{2} - V(\phi),
\end{equation}
from which the inflaton equation-of-state parameter is defined as
\begin{equation}
\omega_\phi = \frac{p_\phi}{\rho_\phi}.
\end{equation}

Using the continuity equation and transforming derivatives via $d/dt = H\,d/dN$, the evolution of the inflaton energy density during reheating is given by
\begin{equation}
    \frac{d\rho_\phi}{dN}
    = -\left(3 + \frac{\Gamma}{H}\right)(1+\omega_\phi)\rho_\phi.
    \label{rhophi}
\end{equation}

Similarly, the radiation energy density evolves as
\begin{equation}
    \frac{d\rho_R}{dN}
    = -4\rho_R + \frac{\Gamma}{H}(1+\omega_\phi)\rho_\phi.
    \label{rhor}
\end{equation}

The effective equation of state during reheating is therefore
\begin{equation}
    \omega_{RH} =
    \frac{\omega_\phi \rho_\phi + \rho_R/3}
         {\rho_\phi + \rho_R}.
\end{equation}

To capture the dynamical behavior of reheating, we numerically integrate Eqs.~\eqref{rhophi} and \eqref{rhor} simultaneously. The time-dependent $\omega_\phi$ during oscillations is obtained by averaging over one oscillation period \cite{Turner}, leading to
\begin{equation}
    \omega_\phi(\rho_\phi)
    =
    2\frac{\int_{\phi_{m1}}^{\phi_{m2}} 
    d\phi\sqrt{1-\frac{V(\phi)}{\rho_\phi}}}
    {\int_{\phi_{m1}}^{\phi_{m2}} 
    d\phi / \sqrt{1-\frac{V(\phi)}{\rho_\phi}}}
    - 1,
    \label{omegaphi}
\end{equation}
where $\phi_{m1}$ and $\phi_{m2}$ denote the turning points of oscillation, determined from $\rho_\phi = V(\phi)$.

For the potential
\begin{equation}
    V(\phi)=V_0\left[\tan^{-1}\left(\frac{\kappa\phi}{M_{pl}}\right)\right]^{2n},
\end{equation}
the oscillation amplitudes are
\begin{equation}
    \phi_{m1,m2}
    = \pm\frac{M_{pl}}{\kappa}
    \tan\left[\left(\frac{\rho_\phi}{V_0}\right)^{1/2n}\right].
\end{equation}

\subsection*{Time-dependent decay rate}

In the simplest treatment, the decay rate $\Gamma$ can be taken as constant. However, in perturbative reheating, the decay rate depends on the inflaton mass, which itself evolves as the field oscillates. This introduces an explicit time dependence in $\Gamma$.

Following an analysis similar to \cite{Garcia}, and assuming that the inflaton decays into fermion pairs, the decay rate becomes
\begin{equation}
    \Gamma_{\phi\to \bar{f}f}(t)
    = \sqrt{2n(2n-1)}
    \frac{y_{eff}^2}{8\pi}
    \left(\frac{V_0\kappa^{2n}}{M_{pl}^4}\right)^{1/2n}
    \left(\frac{\rho_\phi}{M_{pl}^4}\right)^{\frac{n-1}{2n}}
    M_{pl}.
    \label{gammat}
\end{equation}

Thus, the transfer of energy from the inflaton condensate to radiation is governed by a dynamically evolving decay rate, which in turn modifies $\omega_{RH}(t)$, the reheating temperature $T_{RH}$, and the number of reheating e-folds $N_{RH}$. This provides a more realistic and self-consistent description of perturbative reheating compared to the constant EOS approximation.

\section{Results} \label{NA}
\subsection*{Bayesian Inference}
\begin{figure}[H]
    \centering
\begin{subfigure}{0.48\textwidth}
    \includegraphics[width=\linewidth]{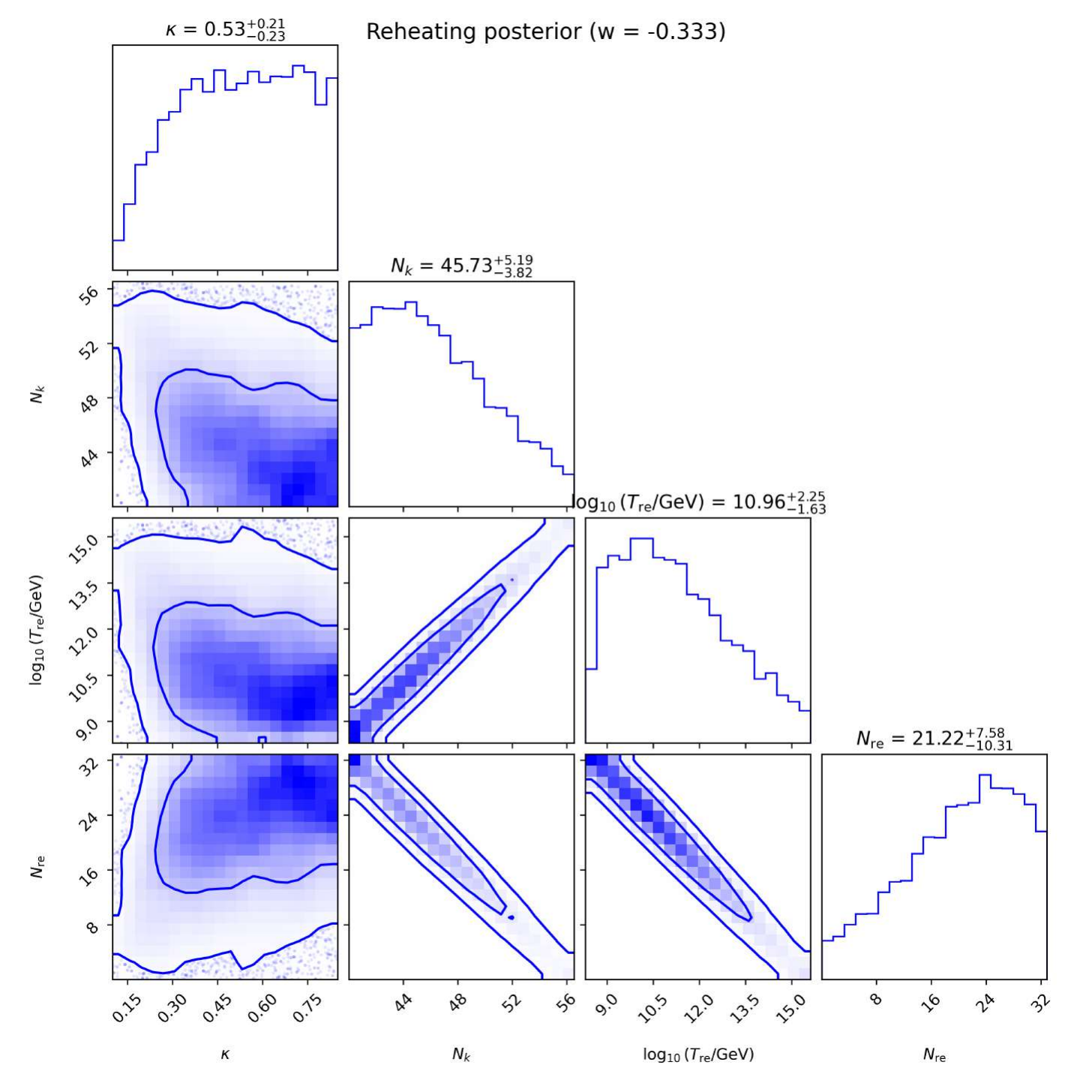}
    
  \end{subfigure}
  \hfill
 \begin{subfigure}{0.48\textwidth}
    \includegraphics[width=\linewidth]{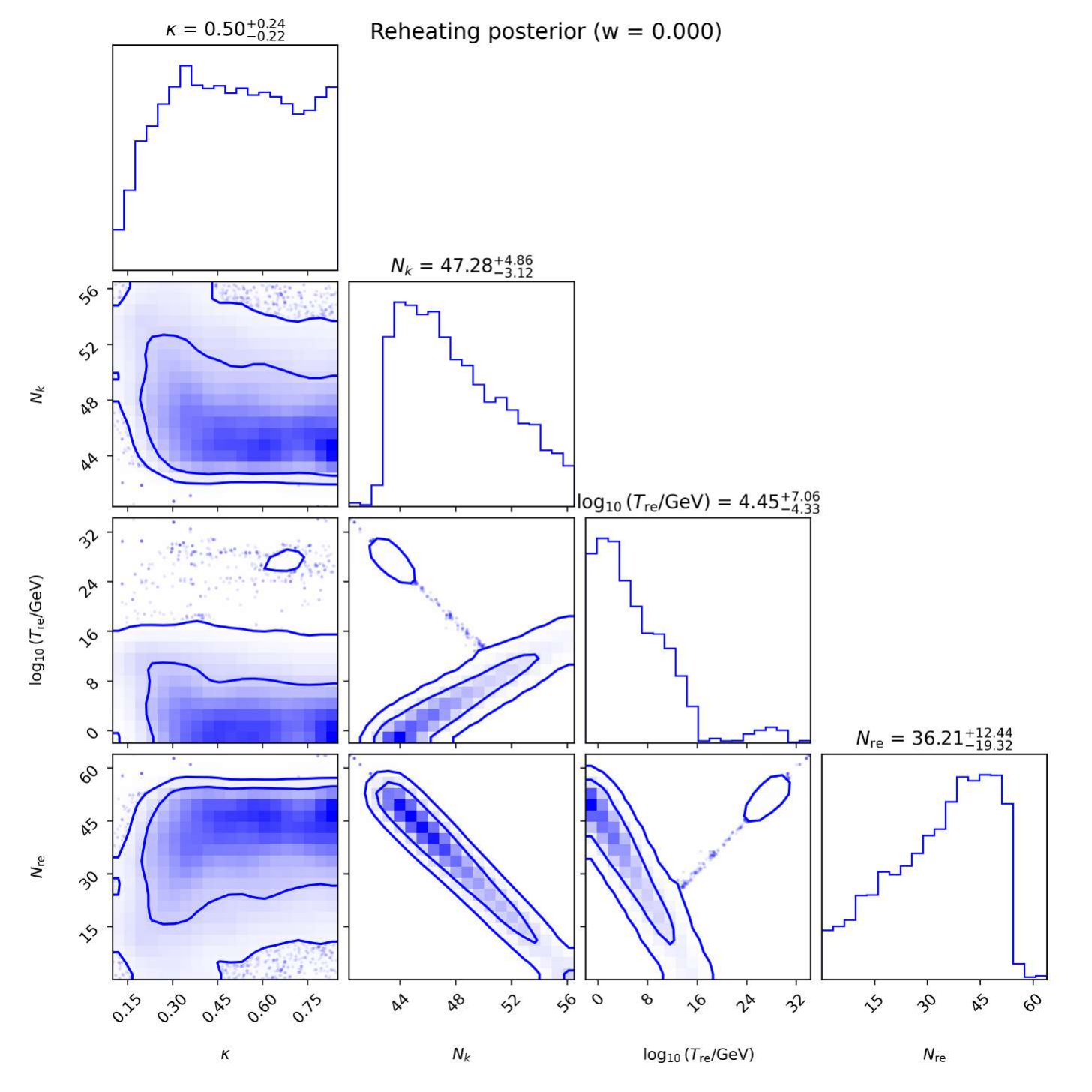}
    
  \end{subfigure}
  \hfill

  \begin{subfigure}{0.48\textwidth}
    \includegraphics[width=\linewidth]{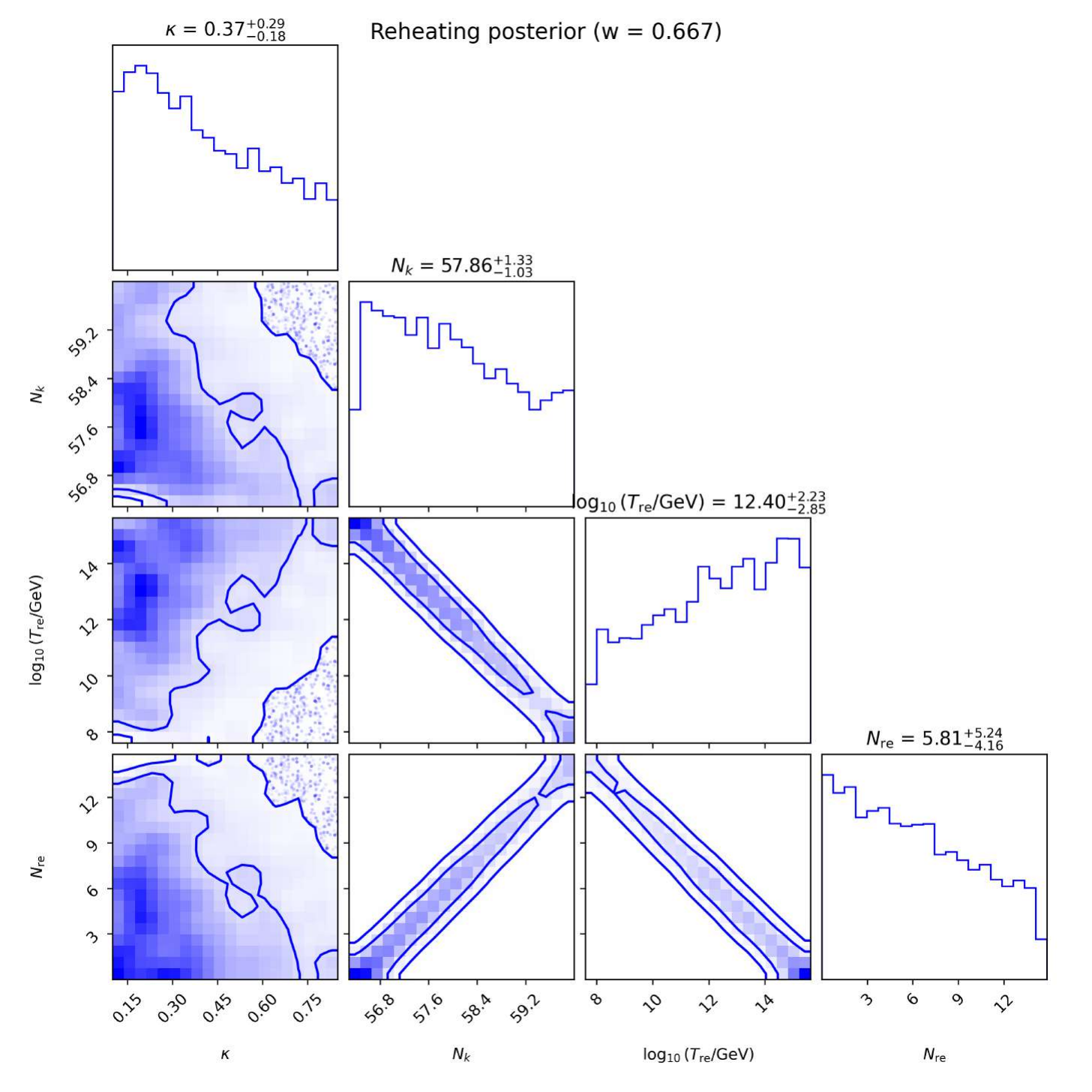}
    
  \end{subfigure}
  \hfill
  \begin{subfigure}{0.48\textwidth}
    \includegraphics[width=\linewidth]{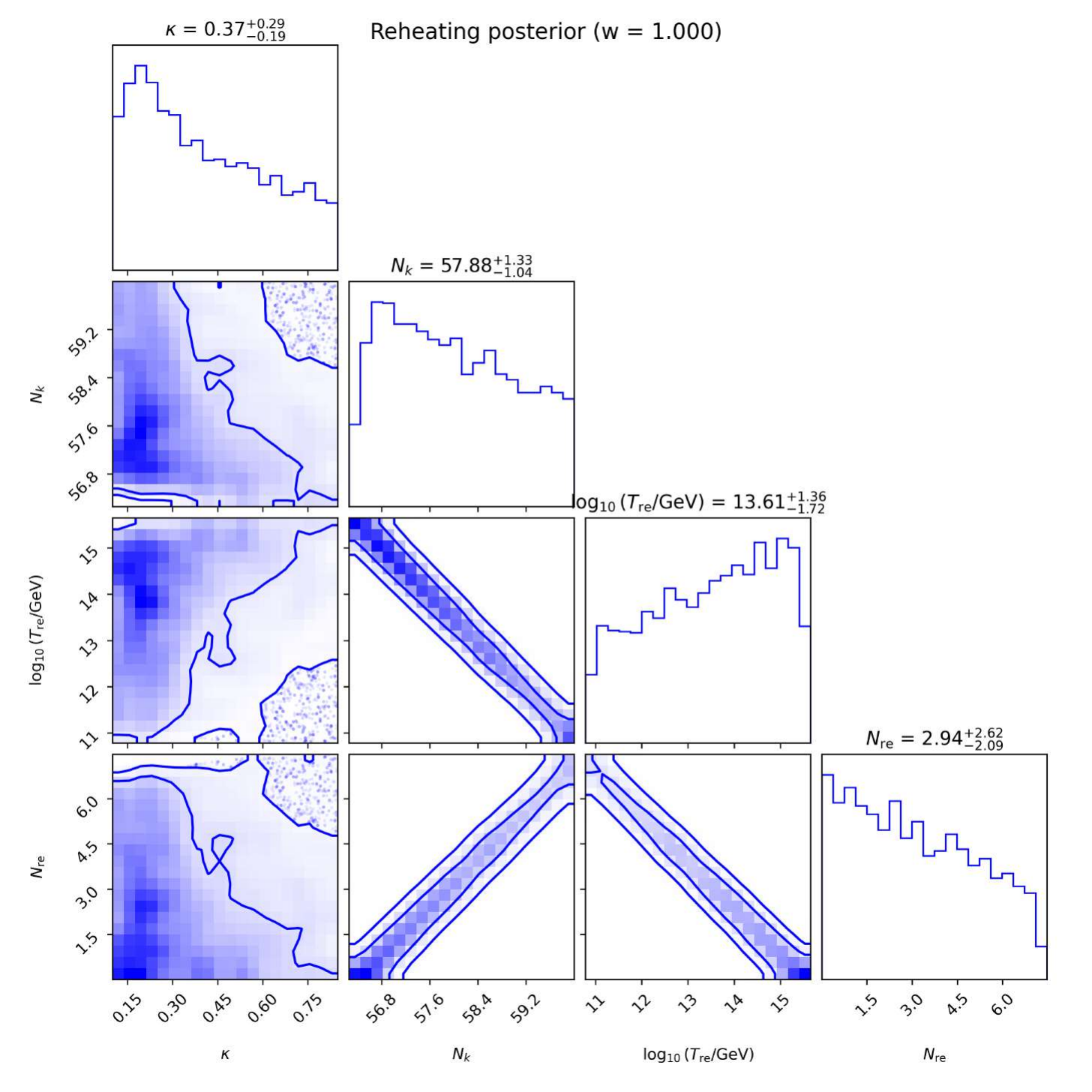}
    
  \end{subfigure}
  \hfill
    \caption{Marginalized posterior distributions and joint confidence contours (dark- $1\sigma$ and light $2\sigma$) bounds for the reheating parameter $\kappa$, number of e-folds during inflation $N_k$, reheating temperature $\log_{10}$($T_{RH}$/GeV) and reheating duration $N_{RH}$ obtained from \emph{Planck} data for fixed reheating EOS, $\omega_{RH}$=[-1/3, 0, 2/3, 1]}
    \label{fig:planck_post}
\end{figure}

\begin{figure}[H]
    \centering
\begin{subfigure}{0.48\textwidth}
    \includegraphics[width=\linewidth]{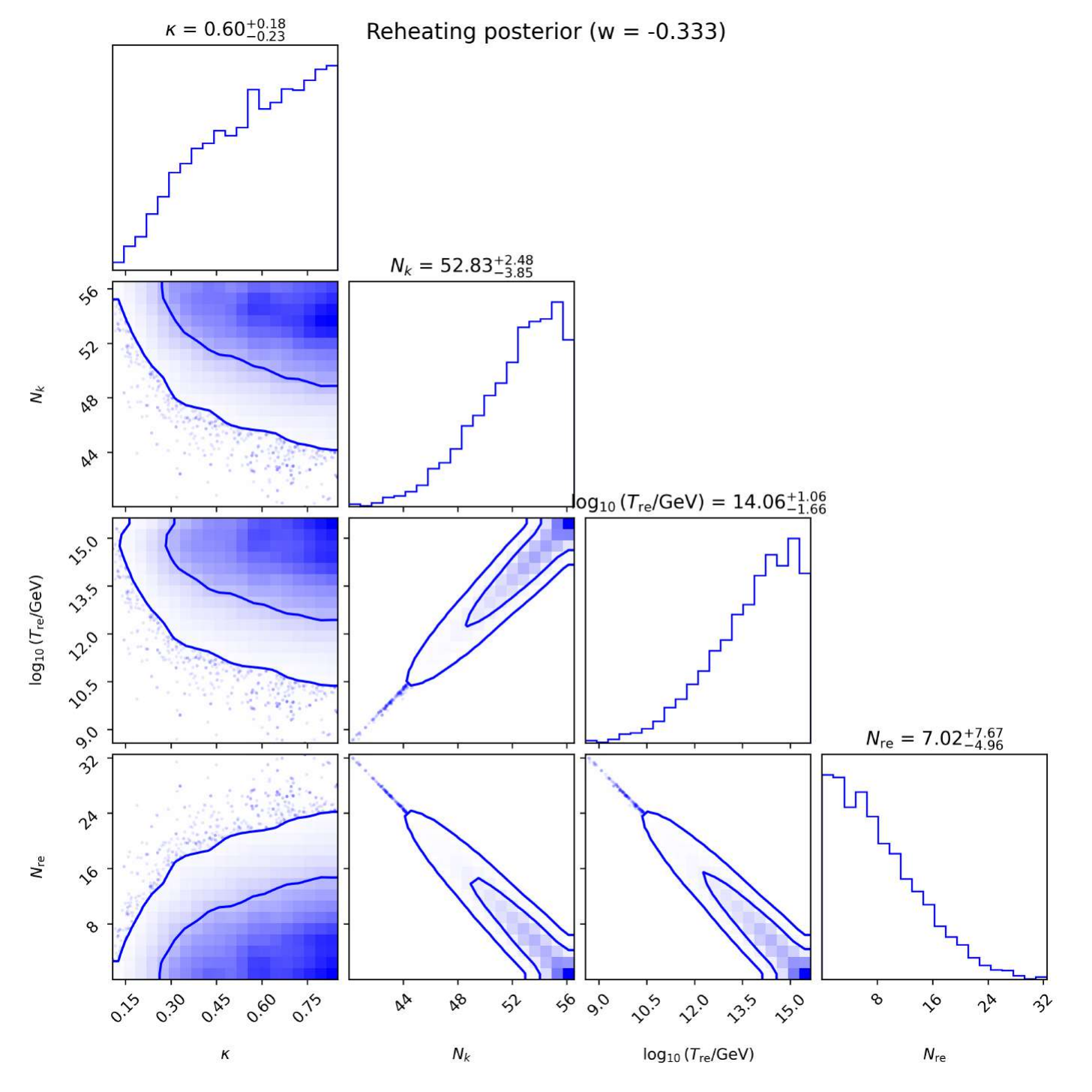}
    
  \end{subfigure}
  \hfill
 \begin{subfigure}{0.48\textwidth}
    \includegraphics[width=\linewidth]{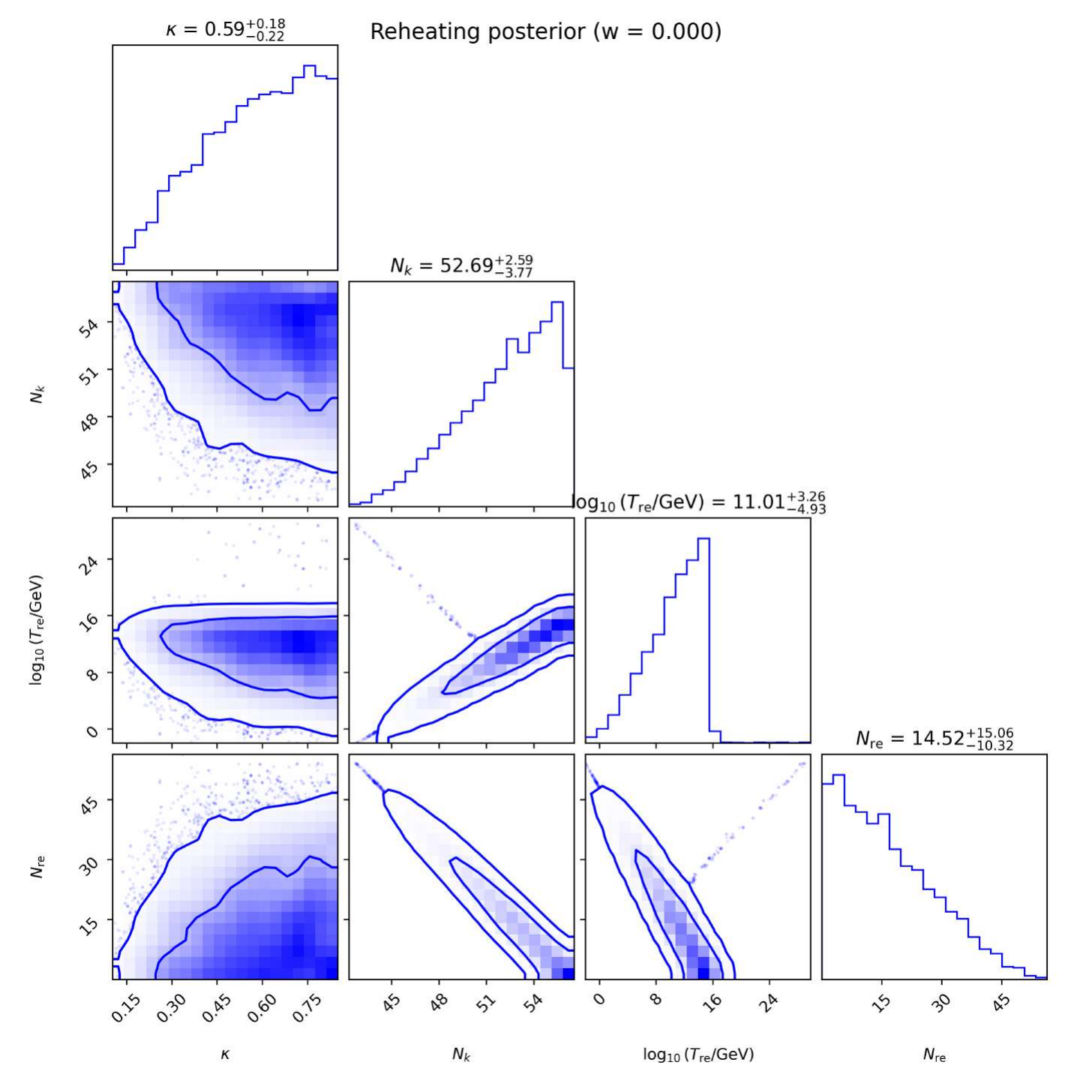}
    
  \end{subfigure}
  \hfill

  \begin{subfigure}{0.48\textwidth}
    \includegraphics[width=\linewidth]{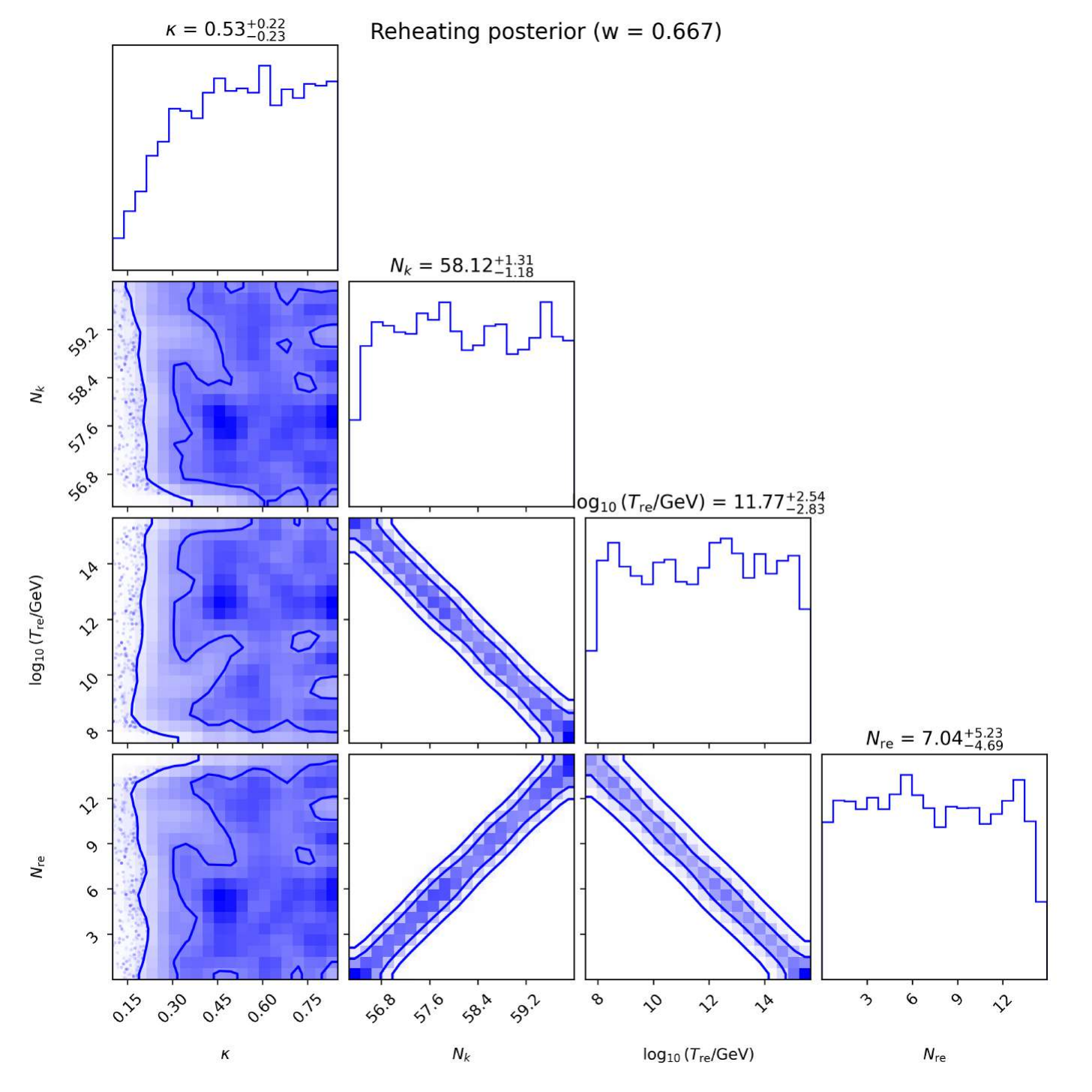}
    
  \end{subfigure}
  \hfill
  \begin{subfigure}{0.48\textwidth}
    \includegraphics[width=\linewidth]{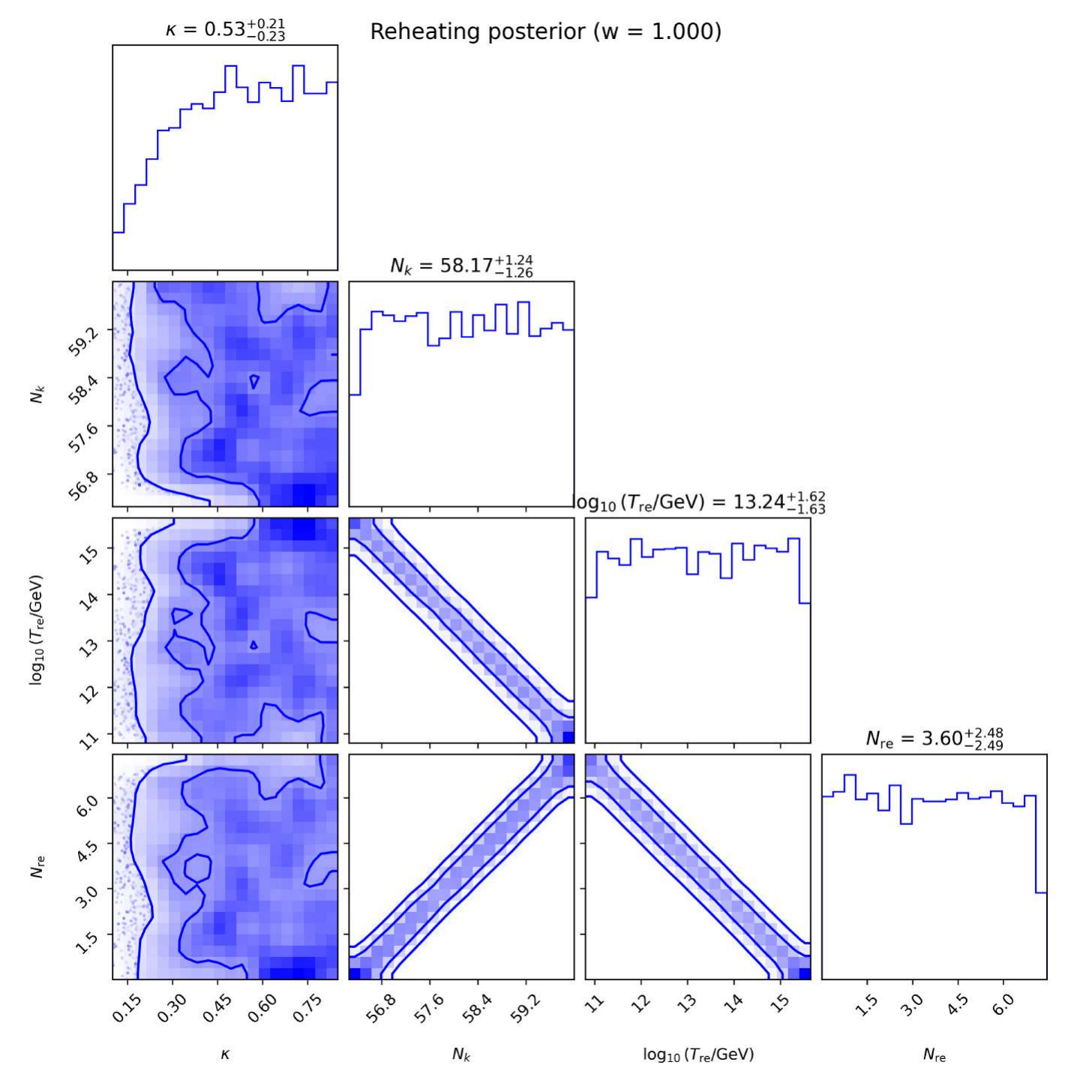}
    
  \end{subfigure}
  \hfill
    \caption{Marginalized posterior distributions and joint confidence contours (dark- $1\sigma$ and light $2\sigma$) bounds for the reheating parameter $\kappa$, number of e-folds during inflation $N_k$, reheating temperature $\log_{10}$($T_{RH}$/GeV) and reheating duration $N_{RH}$ obtained from \emph{ACT} data for fixed reheating EOS, $\omega_{RH}$=[-1/3, 0, 2/3, 1]}
    \label{fig:ACT_post}
\end{figure}

Figures \ref{fig:planck_post} and \ref{fig:ACT_post} show the marginalized posterior distributions and joint confidence contours of the inflationary and reheating parameters ($\kappa$, $N_k$, $\log_{10}(T_{RH}/$GeV),$N_{RH}$) obtained from \emph{Planck} and \emph{ACT} data, respectively, for a fixed reheating equation of state. The reheating parameters are not sampled independently but rather are derived through reheating consistency relations.  

For $\omega_{\rm RH} = -1/3$ and $0$, both \emph{Planck} and \emph{ACT} datasets favor inflationary parameter values in the range $\kappa \sim 0.5$--$0.6$. The posterior width remains relatively broad, with uncertainties of order $\pm 0.2$, indicating that for low reheating equation-of-state values the inflationary sector retains significant freedom due to the extended post-inflationary expansion history.

For these same values of $\omega_{\rm RH}$, \emph{Planck} prefers
\begin{align}
N_k &= 45.73^{+5.19}_{-3.82} \quad (\omega_{\rm RH} = -1/3), \\
N_k &= 47.28^{+4.86}_{-3.12} \quad (\omega_{\rm RH} = 0),
\end{align}
whereas \emph{ACT} systematically favors larger e-folding numbers,
\begin{align}
N_k &= 52.83^{+2.48}_{-3.85} \quad (\omega_{\rm RH} = -1/3), \\
N_k &= 52.69^{+2.59}_{-3.77} \quad (\omega_{\rm RH} = 0).
\end{align}
This shift reflects the slightly different spectral index constraints from \emph{ACT}, which translate into a higher pivot-scale matching requirement and hence a larger preferred $N_k$.

As the reheating equation-of-state parameter increases to $\omega_{\rm RH} = 2/3$ and $1$, a significant shift in the inflationary e-folding number is observed, with both datasets yielding
\begin{align}
N_k &= 57.86^{+1.33}_{-1.03} \quad (\omega_{\rm RH} = 2/3)~(Planck), \\
N_k &= 57.88^{+1.33}_{-1.04} \quad (\omega_{\rm RH} = 1)~(Planck)\\
N_k &= 58.12^{+1.31}_{-1.18} \quad (\omega_{\rm RH} = 2/3)~(ACT), \\
N_k &= 58.17^{+1.24}_{-1.26} \quad (\omega_{\rm RH} = 1)~(ACT).
\end{align}
Notably, the uncertainties shrink to the percent level. This tightening of constraints arises because higher $\omega_{\rm RH}$ leads to a more rapidly redshifting reheating phase, reducing the allowed reheating duration and thereby diminishing degeneracies with $N_k$.

In the reheating sector, \emph{Planck} favors relatively long reheating durations for $\omega_{\rm RH} = -1/3$ and $0$,
\begin{align}
N_{\rm RH} &= 21.22^{+7.58}_{-10.31} \quad (\omega_{\rm RH} = -1/3), \\
N_{\rm RH} &= 36.21^{+12.44}_{-19.32} \quad (\omega_{\rm RH} = 0),
\end{align}
accompanied by comparatively lower reheating temperatures
\begin{align}
T_{\rm RH} &\sim 10^{10.96}~{\rm GeV} \quad (\omega_{\rm RH} = -1/3), \\
T_{\rm RH} &\sim 10^{4.45}~{\rm GeV} \quad (\omega_{\rm RH} = 0).
\end{align}
For higher $\omega_{\rm RH}$, the reheating duration decreases significantly,
\begin{align}
N_{\rm RH} &= 5.81^{+5.24}_{-4.16} \quad (\omega_{\rm RH} = 2/3), \\
N_{\rm RH} &= 2.94^{+2.62}_{-2.09} \quad (\omega_{\rm RH} = 1),
\end{align}
with corresponding increases in reheating temperature,
\begin{align}
T_{\rm RH} &\sim 10^{12.40}~{\rm GeV} \quad (\omega_{\rm RH} = 2/3), \\
T_{\rm RH} &\sim 10^{13.61}~{\rm GeV} \quad (\omega_{\rm RH} = 1).
\end{align}

A qualitatively similar trend is observed with \emph{ACT} data,
\begin{align}
N_{\rm RH} &= 7.02^{+7.67}_{-4.96} \quad (\omega_{\rm RH} = -1/3), \\
N_{\rm RH} &= 14.52^{+15.06}_{-10.32} \quad (\omega_{\rm RH} = 0), \\
N_{\rm RH} &= 7.04^{+5.23}_{-4.69} \quad (\omega_{\rm RH} = 2/3), \\
N_{\rm RH} &= 3.60^{+2.48}_{-2.49} \quad (\omega_{\rm RH} = 1),
\end{align}
with corresponding reheating temperatures
\begin{align}
T_{\rm RH} &\sim 10^{14.06}~{\rm GeV} \quad (\omega_{\rm RH} = -1/3), \\
T_{\rm RH} &\sim 10^{11.01}~{\rm GeV} \quad (\omega_{\rm RH} = 0), \\
T_{\rm RH} &\sim 10^{11.77}~{\rm GeV} \quad (\omega_{\rm RH} = 2/3), \\
T_{\rm RH} &\sim 10^{13.24}~{\rm GeV} \quad (\omega_{\rm RH} = 1).
\end{align}

Overall, both datasets consistently exhibit the expected anti-correlation between reheating duration and temperature, and demonstrate that increasing $\omega_{\rm RH}$ drives the Universe toward progressively shorter and more efficient reheating scenarios. While \emph{ACT} systematically favors slightly larger $N_k$ and higher reheating temperatures compared to \emph{Planck}, the qualitative trends and degeneracy structures remain robust across datasets.

\begin{figure}[H]
    \centering
\begin{subfigure}{0.48\textwidth}
    \includegraphics[width=\linewidth]{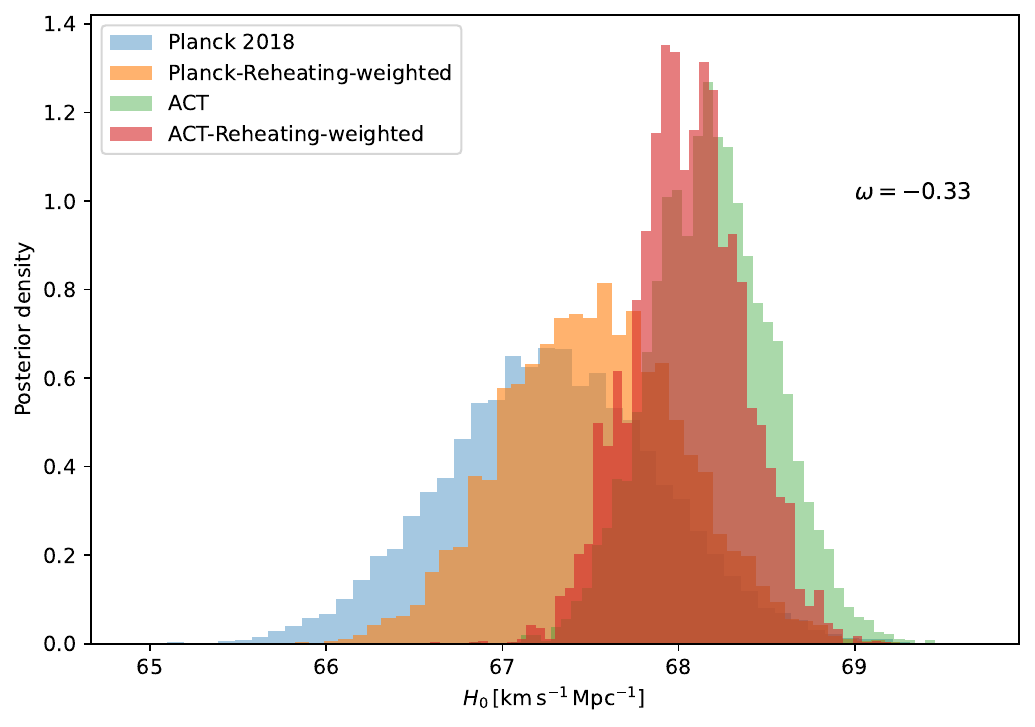}
    \end{subfigure}
  \hfill
 \begin{subfigure}{0.48\textwidth}
    \includegraphics[width=\linewidth]{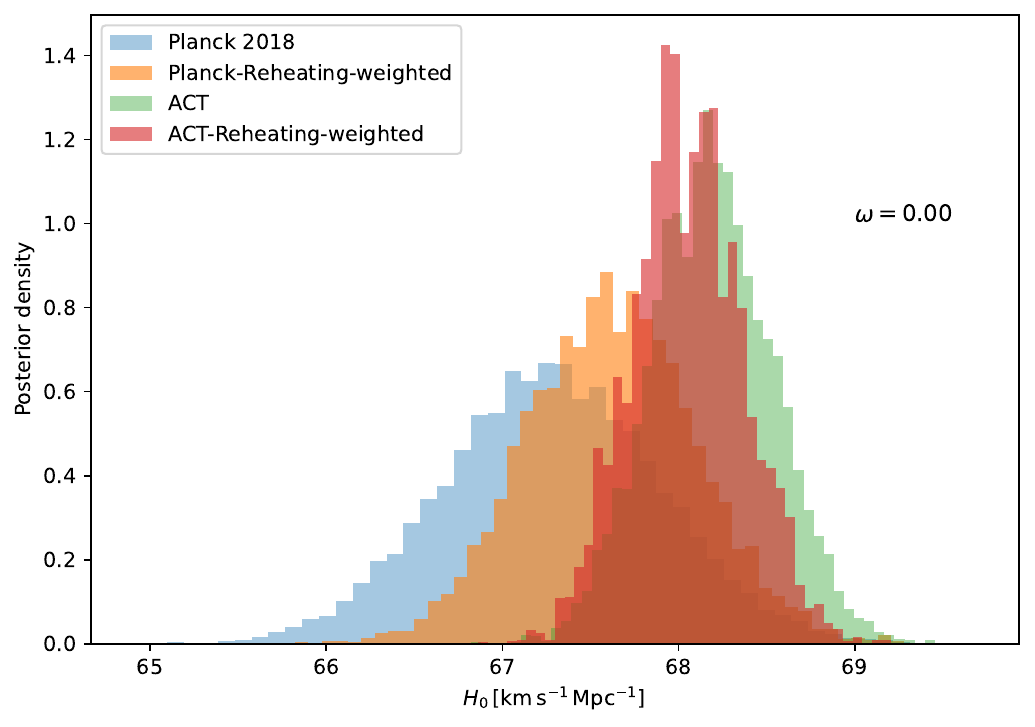}
    \end{subfigure}
  \hfill
\begin{subfigure}{0.48\textwidth}
    \includegraphics[width=\linewidth]{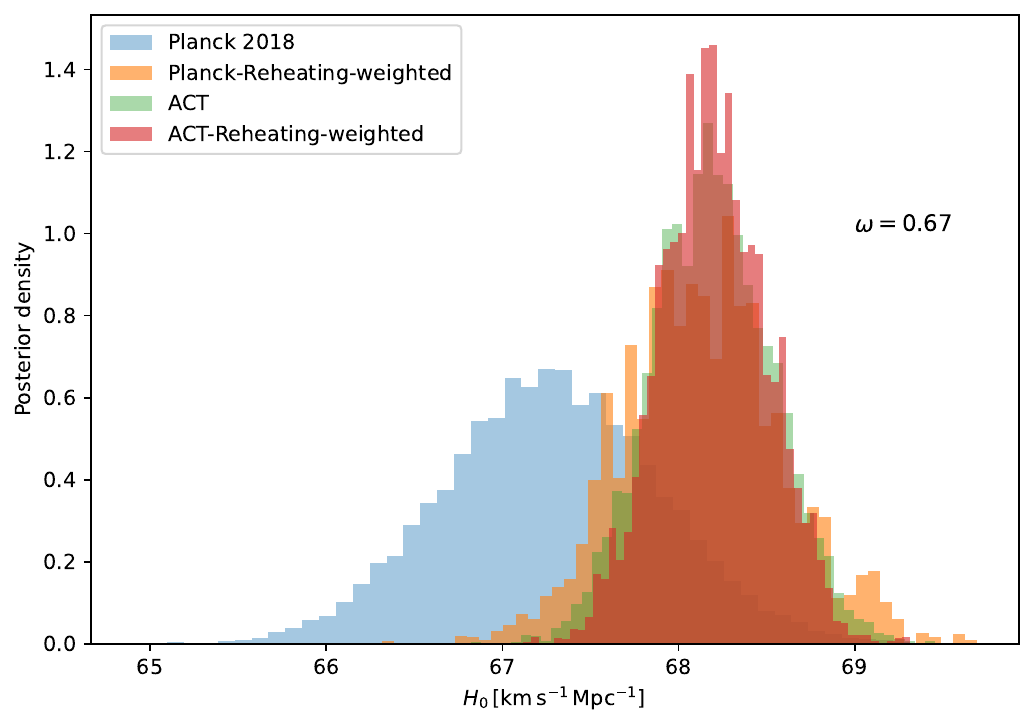}
    \end{subfigure}
  \hfill
  \begin{subfigure}{0.48\textwidth}
    \includegraphics[width=\linewidth]{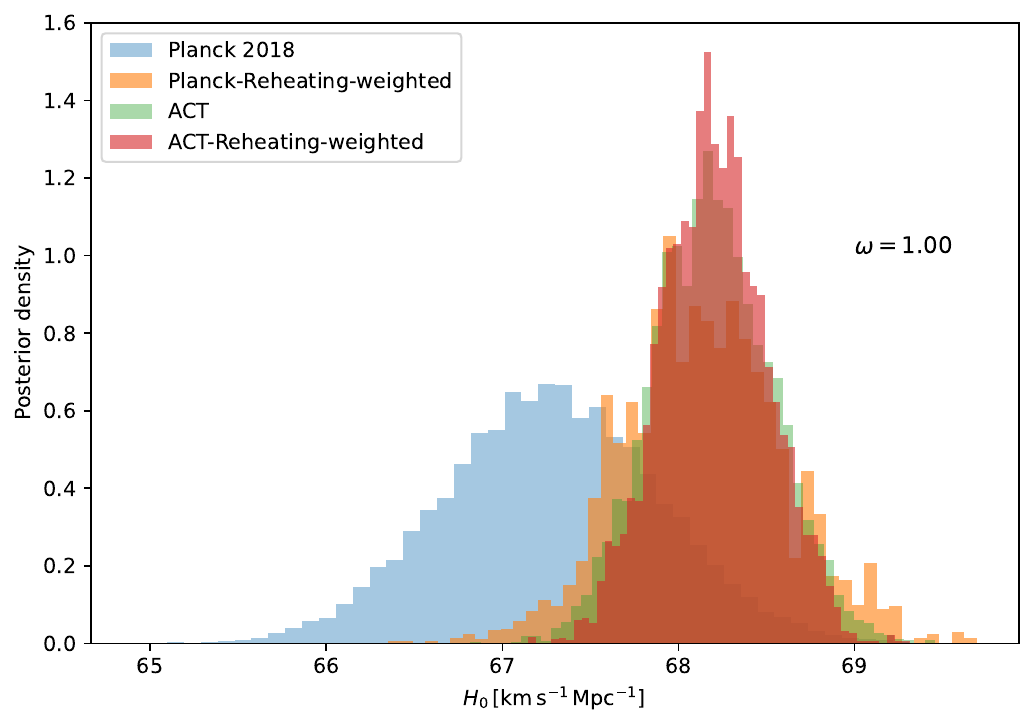}
    \end{subfigure}
  \hfill
    \caption{Posterior distribution of Hubble parameter $H_0$: The blue and green histograms show the marginalized \emph{Planck} and \emph{ACT} posteriors, respectively. The orange and red histograms correspond to the reheating-weighted posteriors}
    \label{ig:placeholder}
\end{figure}

The reheating weighted $H_0$ distributions are obtained via Monte Carlo evaluation of $P(H_0|\mathcal{M})=\int P(H_0|n_s,\mathcal{D})P(n_s|\mathcal{M})dn_s$, where $P(n_s|\mathcal{M})$ is the posterior induced by the inflationary-reheating model ($\mathcal{M}=$ reheating model).   The conditional distribution $P(H_0|n_s,\mathcal{D})$ ($\mathcal{D}=$CMB data) is reconstructed directly from the publicly available MCMC chains (\emph{Planck} and \emph{ACT}) by binning samples in $n_s$ and recording the weighted distribution of $H_0$ within each bin. The reheating dynamics of the model predicts a distribution $P(n_s|\mathcal{M})$ through the relation between number of e-folds, the reheating equation of state, and the inflationary potential. Then the above integral is evaluated via Monte-Carlo sampling by drawing $H_0$ values from the empirical conditional distribution corresponding to each $n_s$ value predicted by the reheating model for different equation of state parameter $\omega_{RH}$. For the case of $\omega_{RH}=-1/3 ~\text{and}~\omega_{RH}=0$ the \emph{Planck} posterior for $H_0$ is clearly separated from the other distributions, peaking at lower values of the Hubble parameter. Once reheating constraints are propagated, however, the \emph{Planck} reheating-weighted posterior shifts toward higher $H_0$ but still remains distinct from the \emph{ACT} posterior. This behavior follows directly from the positive degeneracy between $n_s$ and $H_0$ in the CMB likelihood. The reheating dynamics restrict the allowed range of $N_k$, thereby selecting a narrower region in $n_s$. Since larger values of $n_s$ correlate with larger $H_0$, the reheating-induced selection naturally shifts the \emph{Planck} posterior toward higher $H_0$ without introducing any modification to late-time cosmology.

In contrast, the \emph{ACT} posterior is only mildly affected by reheating weighting. \emph{ACT} already prefers comparatively larger values of $n_s$ and exhibits a steeper $n_s$–$H_0$ degeneracy than \emph{Planck}. As a result, the reheating-preferred region in $n_s$ lies close to the \emph{ACT} maximum-likelihood region, leading to strong overlap between the \emph{ACT} and \emph{ACT}-reheating posteriors. Reheating dynamics, therefore, act primarily to move the \emph{Planck} inference toward the \emph{ACT}-favored region of parameter space, while leaving the \emph{ACT} constraints largely unchanged. For $\omega_{RH}=2/3~\text{and}~\omega_{RH}=1$ even the \emph{Planck} reheating-weighted posterior overlaps to a large extent with the \emph{ACT} and \emph{ACT} reheating-weighted posterior. This is because these chosen reheating histories with higher EOS parameter  prefer higher values of $n_s$.

Importantly, no modifications to late-time cosmology are introduced. The shift in $H_0$ arises purely from propagating early-universe reheating constraints through existing CMB posterior structure. This demonstrates that early-universe reheating physics can influence late-time parameter inference purely through the geometry of the CMB likelihood, and may partially alleviate the internal tension between CMB datasets without invoking additional late-time degrees of freedom.

\subsection*{DEOS Reheating}
\begin{figure}[H]
    \centering
\begin{subfigure}{0.48\textwidth}
    \includegraphics[width=\linewidth]{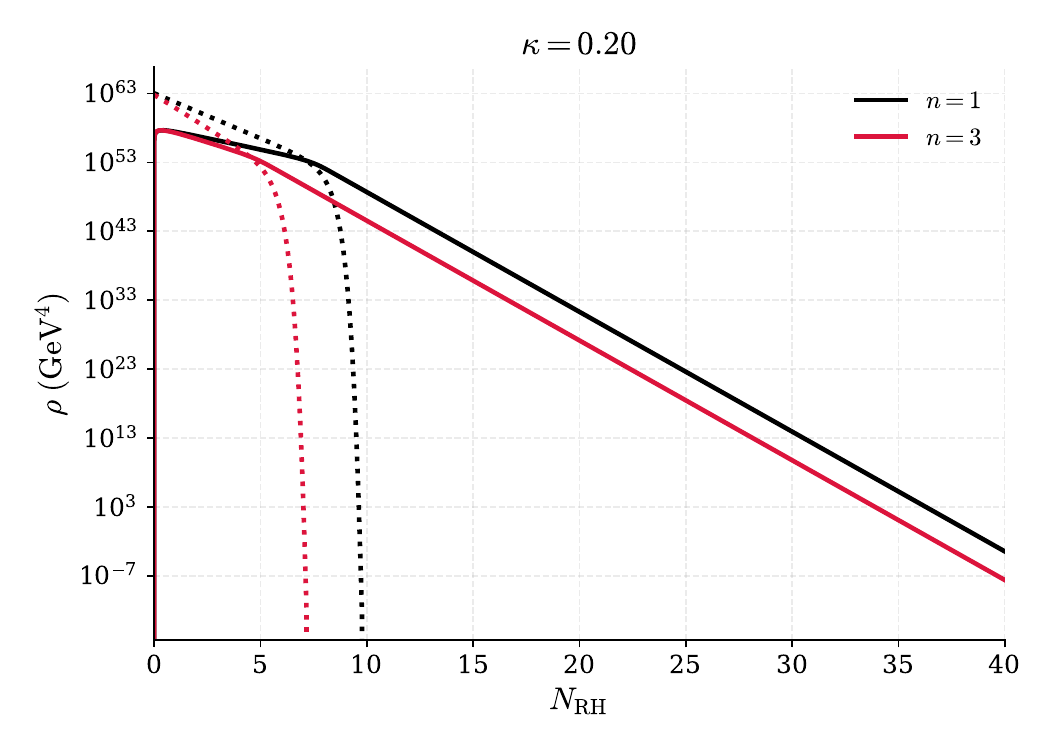}
    
  \end{subfigure}
  \hfill
 \begin{subfigure}{0.48\textwidth}
    \includegraphics[width=\linewidth]{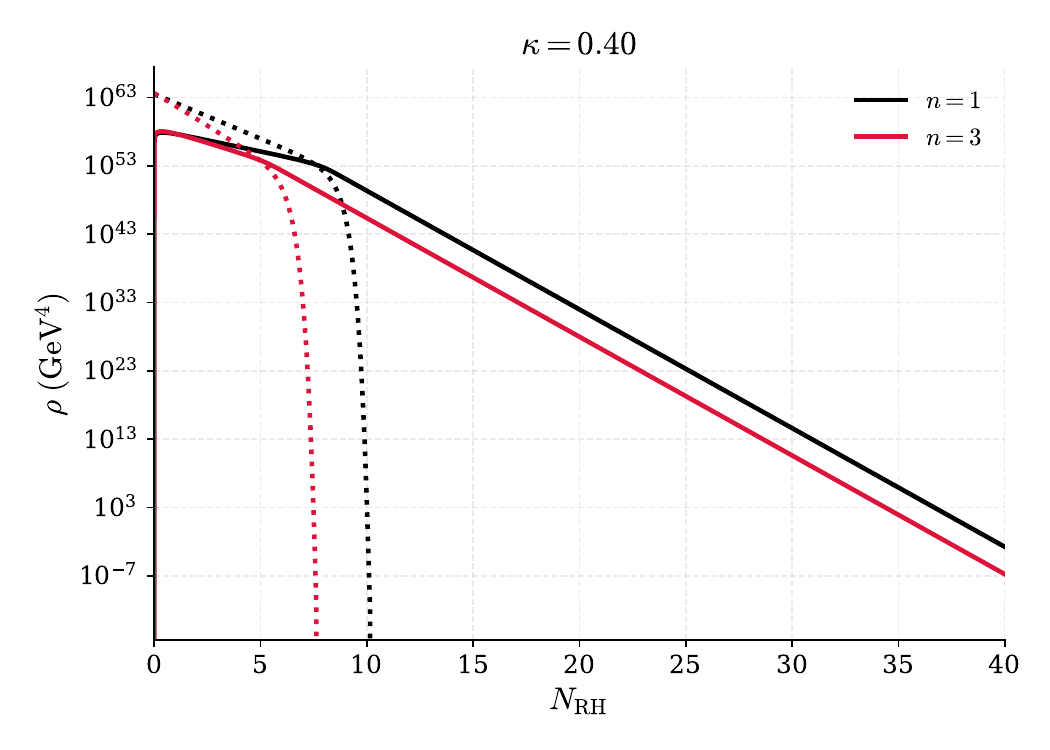}
    
  \end{subfigure}
  \hfill

  \begin{subfigure}{0.48\textwidth}
    \includegraphics[width=\linewidth]{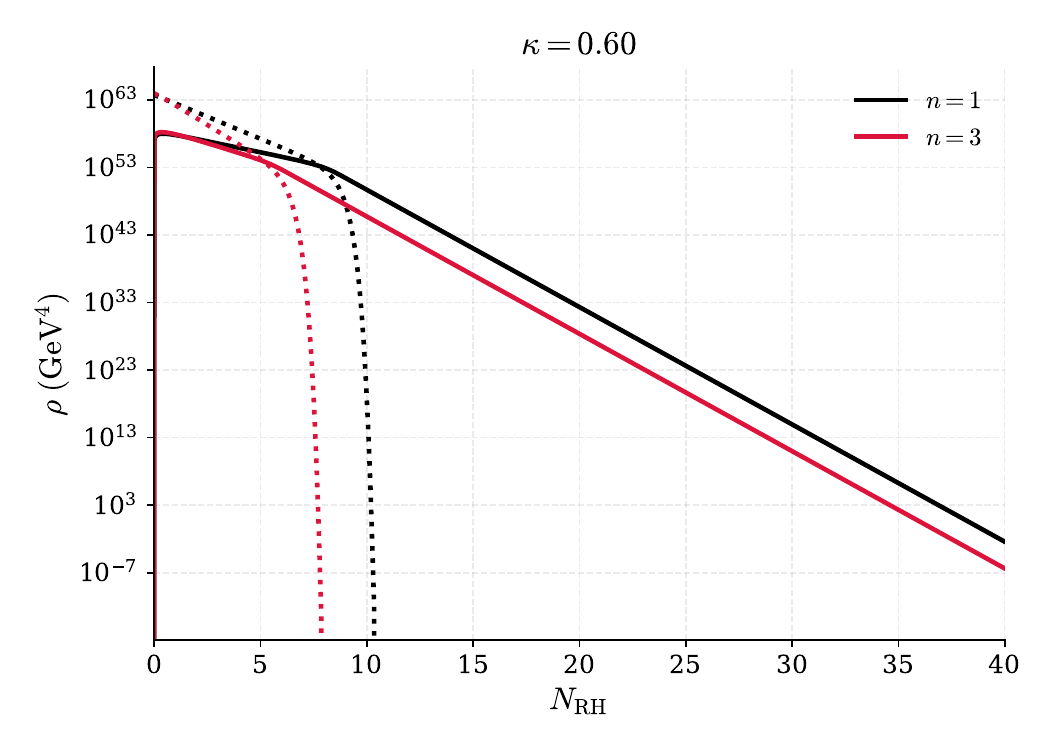}
    
  \end{subfigure}
  \hfill
  \begin{subfigure}{0.48\textwidth}
    \includegraphics[width=\linewidth]{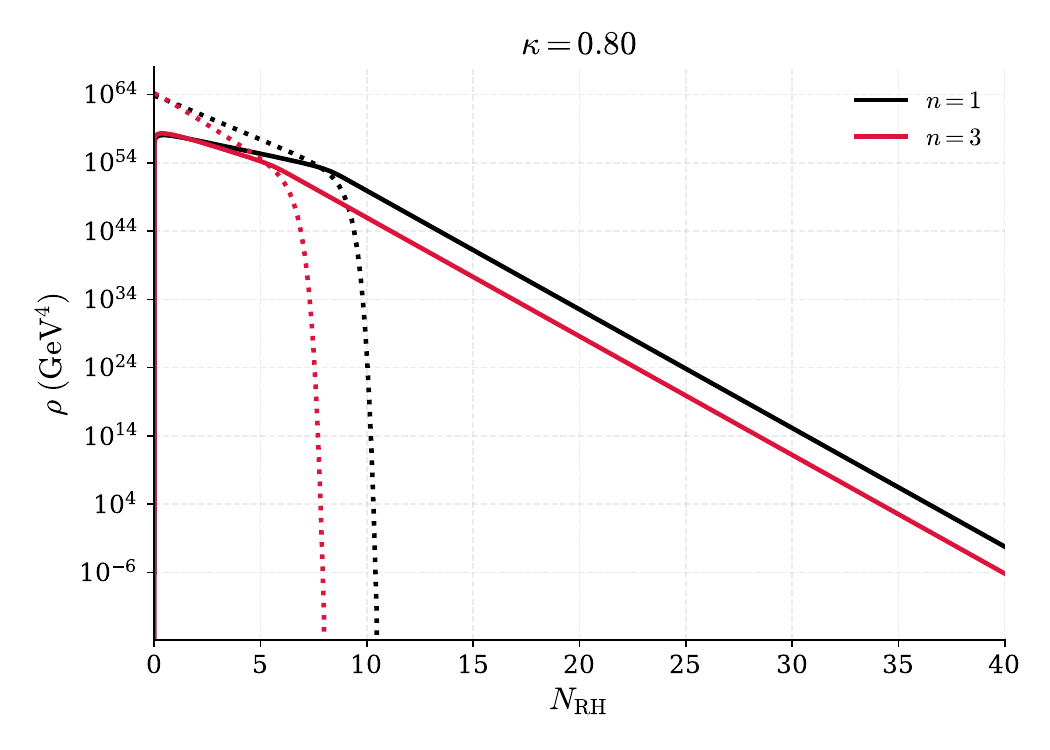}
    
  \end{subfigure}
  \hfill
    \caption{Evolution of energy density with the number of e-folds $N_{RH}$, for $n=1$ (black) and $n=3$ (red) and constant decay rate $\Gamma=10^{-10} {M_{pl}}$. The solid lines correspond to radiation energy density, $\rho_{R}$ and dotted lines correspond to inflaton density $\rho_\phi$.  }
    \label{fig:rho_N}
\end{figure}

\begin{figure}[H]
    \centering
\begin{subfigure}{0.48\textwidth}
    \includegraphics[width=\linewidth]{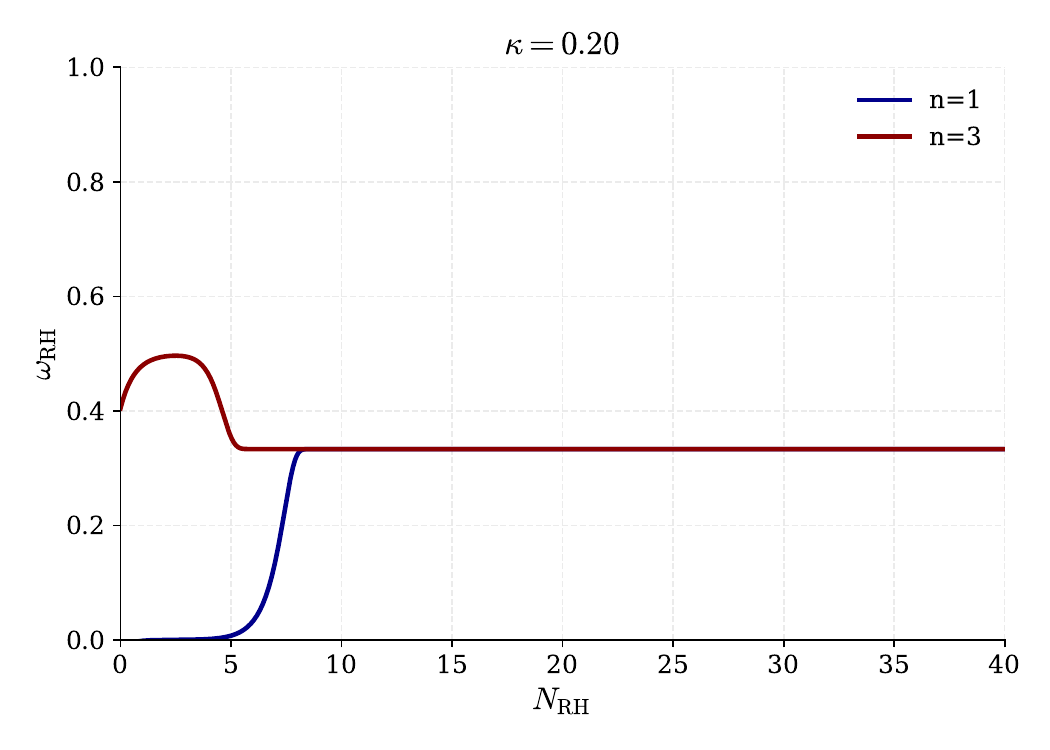}
    
  \end{subfigure}
  \hfill
 \begin{subfigure}{0.48\textwidth}
    \includegraphics[width=\linewidth]{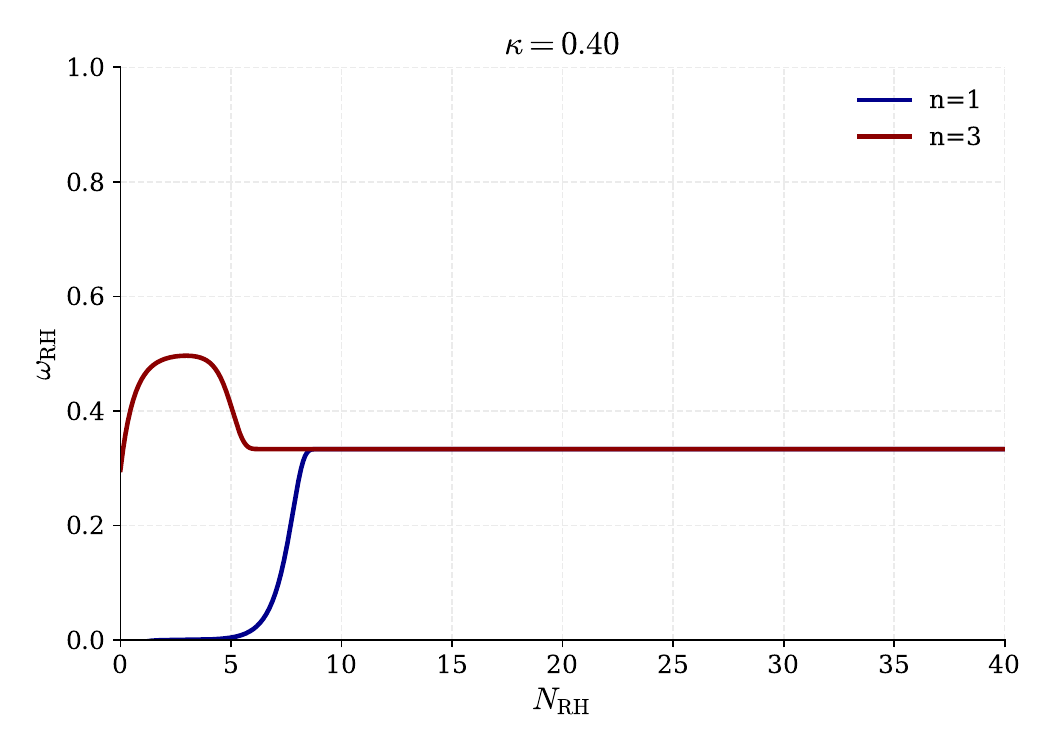}
    
  \end{subfigure}
  \hfill

  \begin{subfigure}{0.48\textwidth}
    \includegraphics[width=\linewidth]{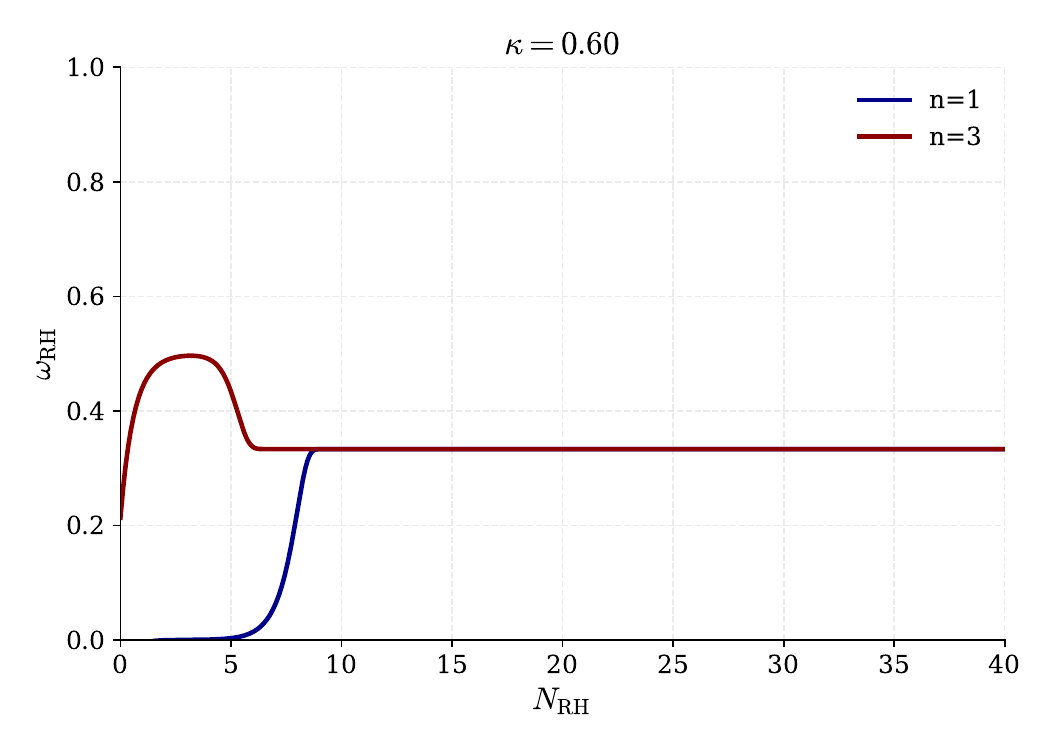}
    
  \end{subfigure}
  \hfill
  \begin{subfigure}{0.48\textwidth}
    \includegraphics[width=\linewidth]{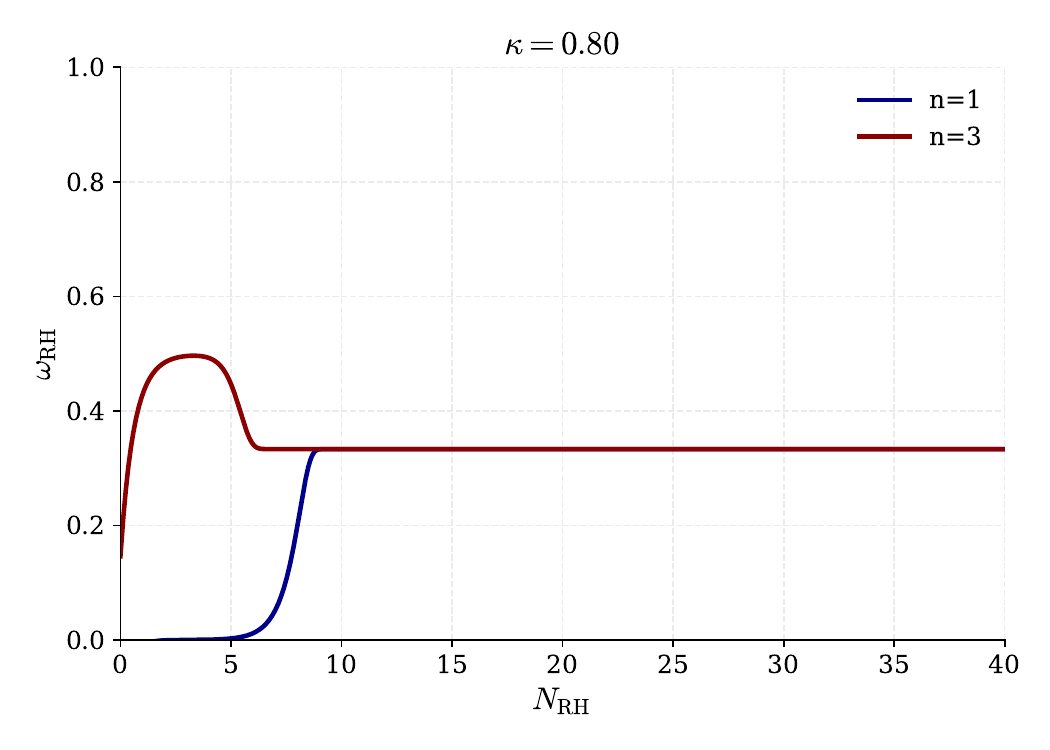}
    
  \end{subfigure}
  \hfill
    \caption{Evolution of EOS parameter $\omega_{RH}$ with the number of e-folds $N_{RH}$, for $n=1$ (black) and $n=3$ (red) and constant decay rate $\Gamma=10^{-10} {M_{pl}}$.}
    \label{fig:omega_N}
\end{figure}

In Figs.~\ref{fig:rho_N} and \ref{fig:omega_N}, we show the evolution of the energy density $\rho_\phi,~\rho_R$ and the effective equation-of-state parameter $\omega$ as functions of the number of e-folds $N_{RH}$, for a constant decay rate $\Gamma$. The evolution clearly demonstrates that reheating is not instantaneous. Instead, the system undergoes a gradual transfer of energy from the inflaton sector to radiation, with inflaton-radiation equality occurring after approximately $7$--$8$ e-folds for $n=1$. The corresponding reheating temperature is of order $T_{\rm RH} \sim 10^{12}$--$10^{13}$~GeV.

Following the equality epoch, the inflaton energy density rapidly becomes subdominant, and the effective equation of state asymptotically approaches $\omega_{RH} = 1/3$, signaling the onset of a radiation-dominated phase. The smooth transition of $\omega$ from its oscillatory inflaton-dominated behavior to the radiation value reflects the continuous nature of the decay process and confirms the consistency of the reheating dynamics within the adopted framework.

A similar analysis is performed for $n=3$. In this case, reheating proceeds more efficiently: inflaton-radiation equality is achieved after only $4$--$5$ e-folds, and the reheating temperature increases to $T_{\rm RH} \sim 10^{13}$~GeV. The faster energy transfer can be understood from the modified inflaton dynamics associated with larger $n$, which alters the oscillatory behavior of the scalar field and enhances the effective dilution of the inflaton energy density. Consequently, the radiation component dominates earlier, shortening the reheating duration.

\renewcommand{\arraystretch}{1.3}

\begin{table}[H]
    \centering
    \begin{subtable}[b]{0.48\textwidth}
\centering
\begin{adjustbox}{max width=\textwidth}
     \addtolength{\tabcolsep}{-0.5pt}
        \small
    \begin{tabular}{|c c c|}
    \hline
    $n=1$ &&\\
    \hline
     $\kappa$ & $N_{RH}$ & $T_{RH}~(GeV)$ \\
     
     \hline
     0.2 & 7.168 & $8.506\times10^{12} $ \\
     
     0.4 & 7.561 & $8.460\times10^{12}$ \\
     
     0.6 & 7.749 & $8.472\times10^{12}$ \\
     
     0.8 & 7.860 & $8.512\times10^{12}$ \\
     
    \hline
\end{tabular}
\end{adjustbox}
\end{subtable}
    \hfill
    \begin{subtable}[b]{0.48\textwidth}
\centering
\begin{adjustbox}{max width=\textwidth}
     \addtolength{\tabcolsep}{-0.5pt}
        \small
    \begin{tabular}{|c c c|}
    \hline
    $n=3$ &&\\
    \hline
     $\kappa$ & $N_{RH}$ & $T_{RH}~(GeV)$ \\
     
     \hline
     0.2 & 4.470 & $1.244\times10^{13} $ \\
     
     0.4 & 4.940 & $1.244\times10^{13}$ \\
     
     0.6 & 5.510 & $1.246\times10^{13}$ \\
     
     0.8 & 5.293 & $1.234\times10^{13}$ \\
     
    \hline
    
    \end{tabular}
    \end{adjustbox}
    \end{subtable}

    \caption{Values of e-folds of reheating $N_{RH}$ and reheating temperature $T_{RH}$ when $\rho_\phi=\rho_R$ for $n=1$ (left) and $n=3$ (right) for constant decay rate $\Gamma=10^{-10} M_{pl}$. }
    \label{tab:para1}
\end{table}

\begin{figure}[H]
    \centering
\begin{subfigure}{0.48\textwidth}
    \includegraphics[width=\linewidth]{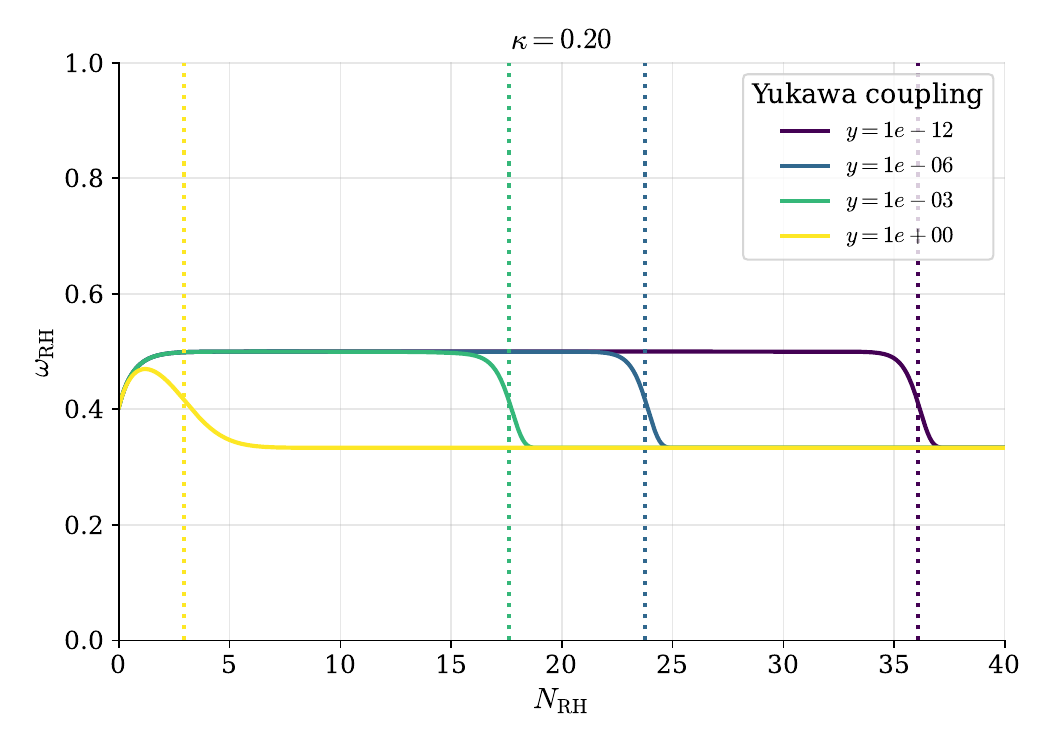}
    \end{subfigure}
  \hfill
 \begin{subfigure}{0.48\textwidth}
    \includegraphics[width=\linewidth]{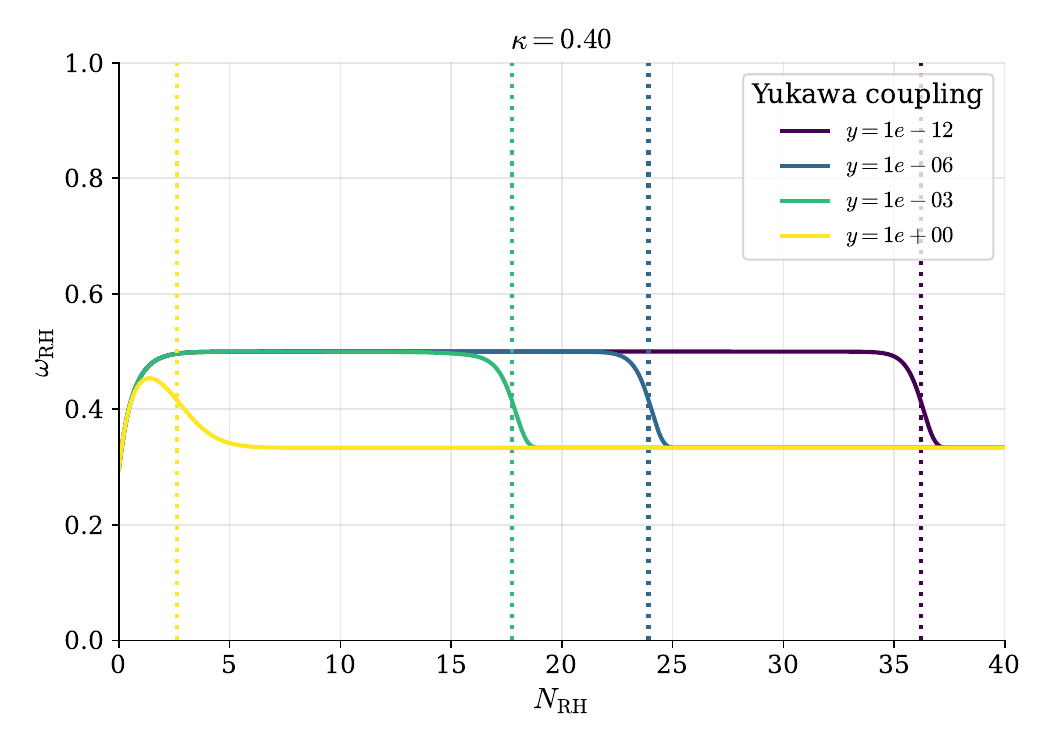}
    \end{subfigure}
  \hfill
\begin{subfigure}{0.48\textwidth}
    \includegraphics[width=\linewidth]{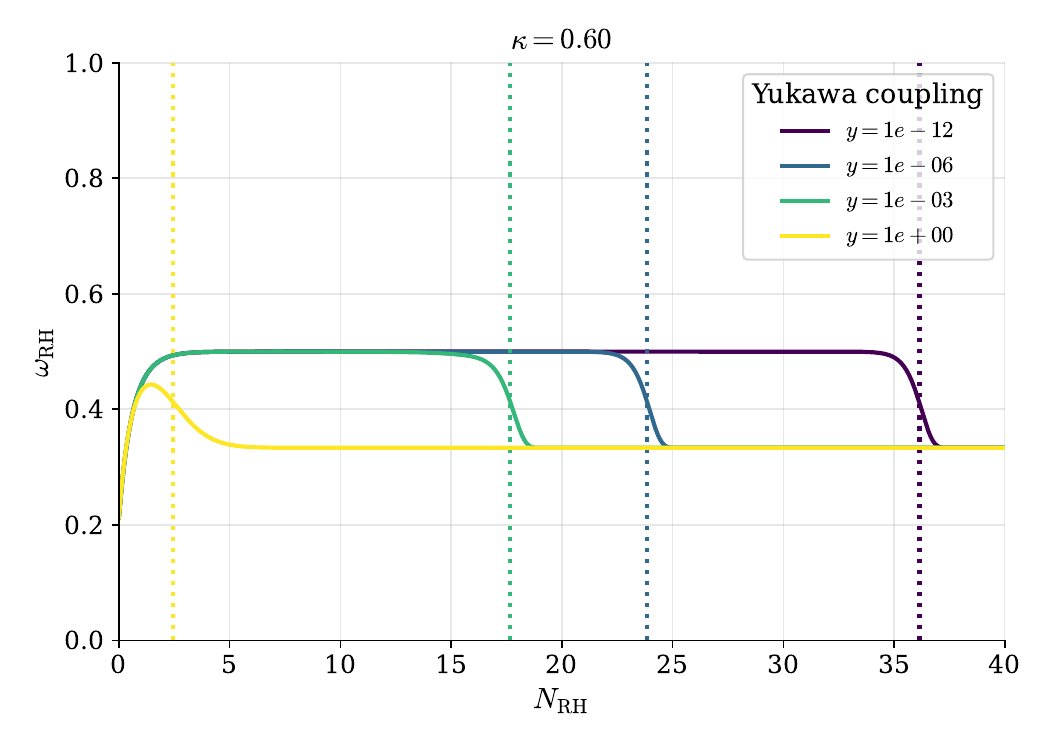}
    \end{subfigure}
  \hfill
  \begin{subfigure}{0.48\textwidth}
    \includegraphics[width=\linewidth]{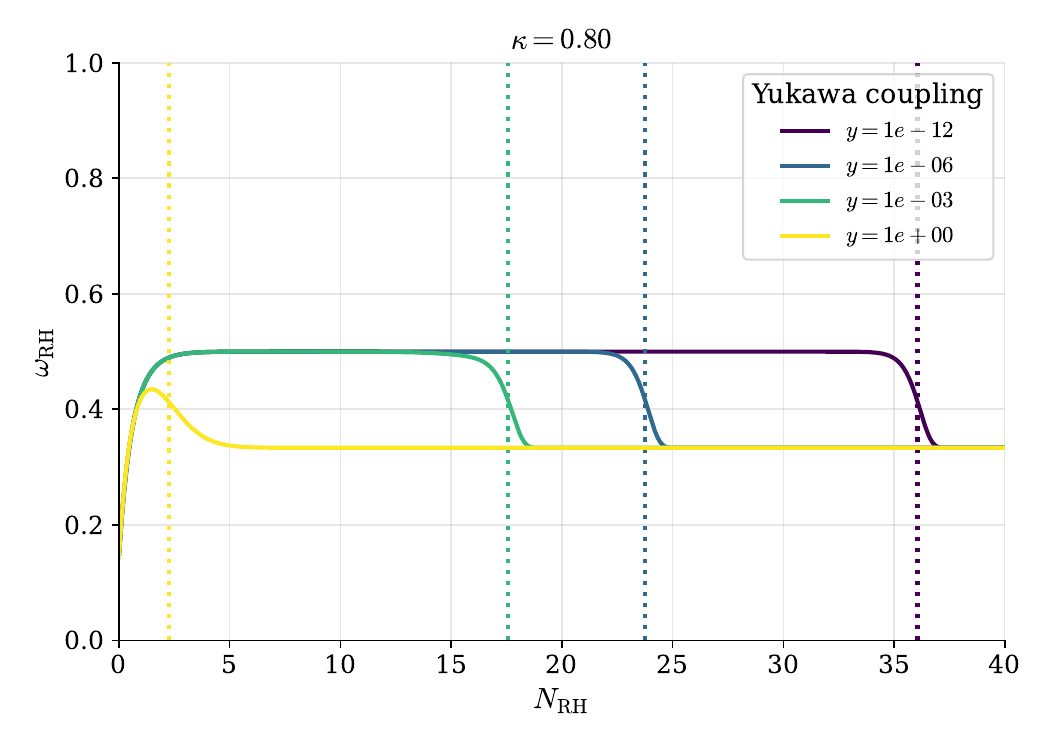}
    \end{subfigure}
  \hfill
    \caption{Evolution of EOS parameter with the number of e-folds $N_{RH}$ for n=3 and time-dependent decay rate $\Gamma(t)$ for different Yukawa couplings $y$.}
    \label{fig:omega_t_N_3_1}
\end{figure}

We next extend the analysis to the case of a dynamic decay rate $\Gamma(t)$ within the DEOS framework. As evident from Eqs.~\eqref{omegaphi} and \eqref{gammat}, non-trivial time dependence in the decay rate arises for $n \geq 2$, motivating a detailed study of the $n=3$ case. The decay rate now depends explicitly on the inflaton dynamics and on the Yukawa coupling $y$, introducing an additional physical handle on the reheating timescale.
\begin{figure}[H]
    \centering
\begin{subfigure}{0.48\textwidth}
    \includegraphics[width=\linewidth]{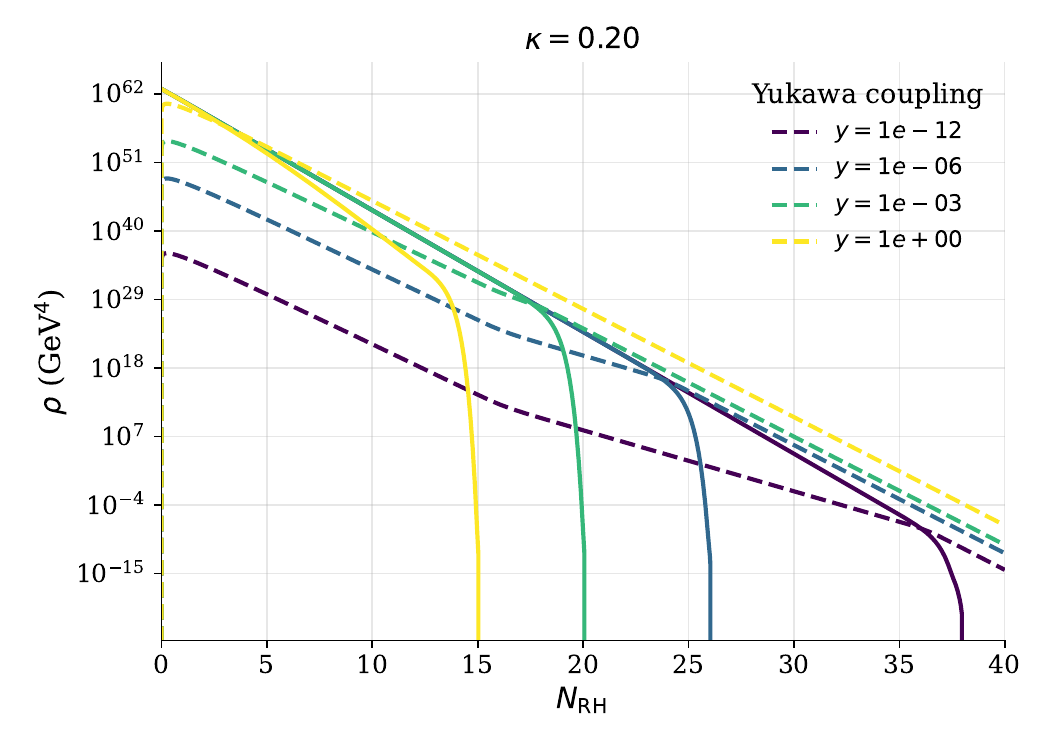}
    \end{subfigure}
  \hfill
 \begin{subfigure}{0.48\textwidth}
    \includegraphics[width=\linewidth]{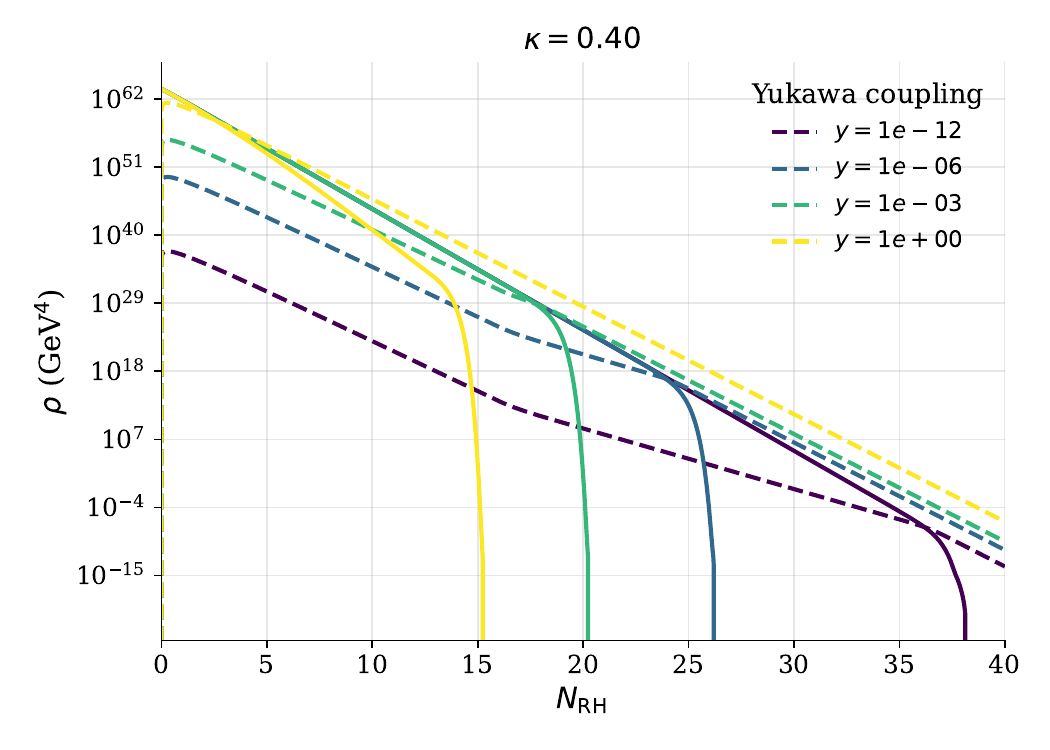}
    \end{subfigure}
  \hfill
\begin{subfigure}{0.48\textwidth}
    \includegraphics[width=\linewidth]{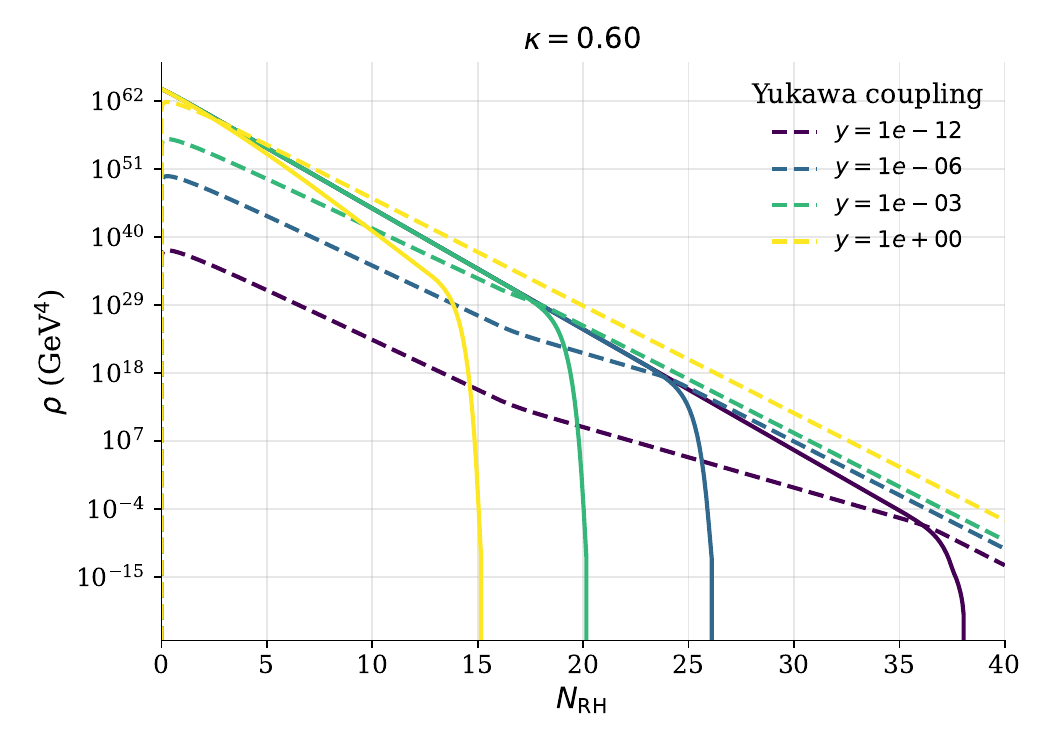}
    \end{subfigure}
  \hfill
  \begin{subfigure}{0.48\textwidth}
    \includegraphics[width=\linewidth]{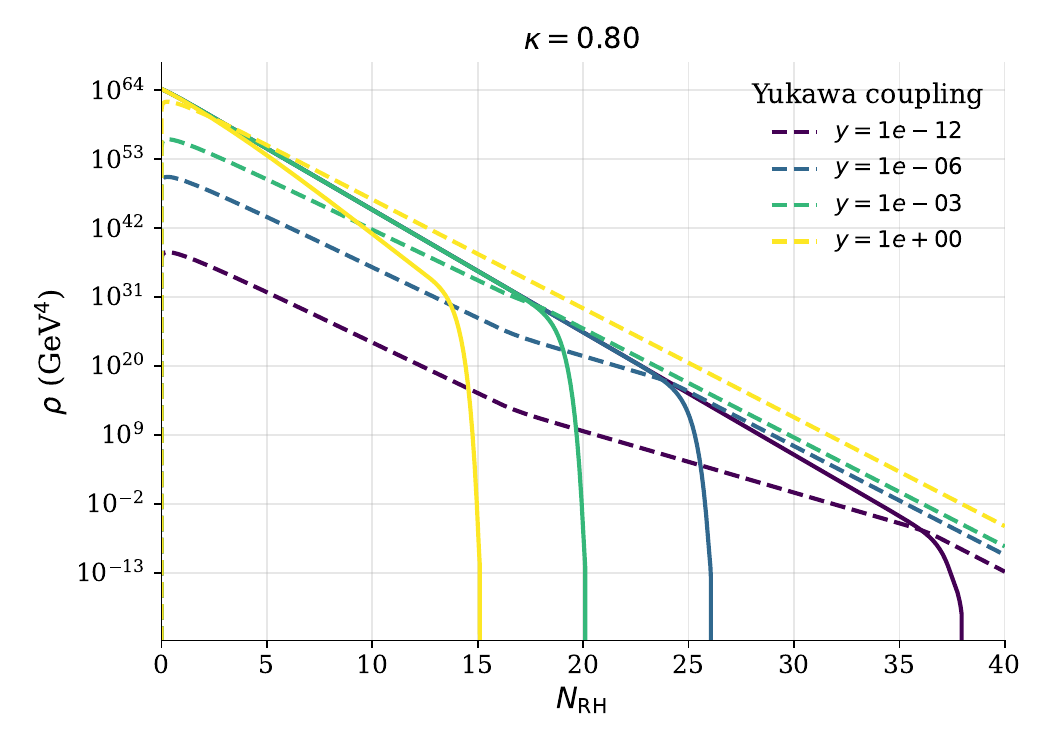}
    \end{subfigure}
  \hfill
    \caption{Evolution of energy density with the number of e-folds $N_{RH}$ for n=3 and time-dependent decay rate for different Yukawa couplings $y$. The solid lines correspond to radiation energy density, $\rho_{R}$ and the dotted lines correspond to inflaton density $\rho_\phi$. }
    \label{fig:rho_t_N_3_1}
\end{figure}
From Fig.\ref {fig:rho_t_N_3_1} and \ref{fig:omega_t_N_3_1}, it is evident that varying the inflationary parameter $\kappa$ in the range $\kappa = 0.2$--$0.8$ produces only negligible changes in the reheating dynamics. The onset of inflaton-radiation equality, the slope of the energy density evolution, and the overall duration of reheating remain essentially unchanged for fixed Yukawa coupling $y$. This behavior can be understood from the reheating equations,
\begin{equation}
\dot{\rho}_\phi + 3H(1+\omega_\phi)\rho_\phi = -\Gamma \rho_\phi,
\end{equation}
where the timescale of energy transfer is governed primarily by the decay rate $\Gamma$. While $\kappa$ controls the inflationary dynamics and the amplitude of the potential during slow-roll, the reheating phase is largely determined by the microphysical decay process. Consequently, the reheating duration scales approximately as $N_{\rm RH} \sim H/\Gamma$ and is therefore much more sensitive to the decay rate than to the precise value of $\kappa$. This explains the near-independence of reheating evolution on $\kappa$ observed in the numerical results.

In contrast, the Yukawa coupling $y$ plays a decisive role in shaping the reheating history. Since the decay rate behaves as $\Gamma(t) \propto y^2 m_\phi(t)$ as can be understood from Eq.~\eqref{gammat}, decreasing $y$ suppresses the inflaton decay and prolongs the reheating phase, whereas larger values of $y$ enhance the decay rate and lead to more rapid energy transfer into radiation. The range of couplings considered, $y=10^{-12}$--$1$, spans physically distinct regimes: extremely feeble portal-like interactions ($y\sim10^{-12}$), Standard-Model–like Yukawa strengths ($y\sim10^{-6}$--$10^{-3}$), and strongly coupled scenarios ($y\sim1$). 

Table~\ref{tab:para} further quantifies the dependence of the reheating duration and temperature on the inflationary parameter $\kappa$ and the Yukawa coupling $y$. For fixed $y$, the reheating e-folds $N_{\rm RH}$ vary only minimally as $\kappa$ is changed from $0.2$ to $0.8$. In particular, for $y=10^{-3}$, $10^{-6}$, and $10^{-12}$, the variation in $N_{\rm RH}$ across the full $\kappa$ range is below $0.2$ e-folds, demonstrating that reheating dynamics are largely insensitive to the precise value of the inflationary parameter. This confirms that $\kappa$ primarily governs the slow-roll phase, while the reheating timescale is controlled by the microphysical decay process.In contrast, the Yukawa coupling $y$ strongly determines the reheating history. As $y$ decreases from $1$ to $10^{-12}$, the reheating duration increases from $\mathcal{O}(1)$ e-folds to $\mathcal{O}(30)$ e-folds. Correspondingly, the reheating temperature exhibits the expected perturbative scaling $T_{\rm RH} \propto \sqrt{\Gamma M_{\rm Pl}} \propto y$, decreasing from $\sim 10^{14}$~GeV for $y=1$ to $\sim 10^{-2}$~GeV for $y=10^{-12}$. The smallest Yukawa coupling $y=10^{-12}$ yields reheating temperatures in the range $T_{RH}\simeq4.6-11.5$ MeV, placing the model close to the lower limit required by successful Big Bang Nucleosynthesis \cite{Salas, Barbieri}. Future improvements in low reheating constraints could therefore disfavor or exclude this regime. The numerical results therefore consistently reproduce the analytic expectations of perturbative reheating within the DEOS framework.

This analysis demonstrates that, within the DEOS scenario, reheating is not characterized by a single fixed timescale but instead emerges dynamically from the interplay between the inflaton potential, the oscillatory equation of state, and the microphysical coupling strength. The resulting dependence of $N_{\rm RH}$ and $T_{\rm RH}$ on both $n$ and $y$ provides a physically motivated origin for the structured degeneracies observed in the Bayesian parameter constraints presented in the previous section.

\begin{table}[H]
    \centering
    \begin{subtable}[b]{0.48\textwidth}
\centering
\begin{adjustbox}{max width=\textwidth}
     \addtolength{\tabcolsep}{-0.5pt}
        \small
    \begin{tabular}{|c c c|}
    \hline
    $y=1$ &&\\
    \hline
     $\kappa$ & $N_{RH}$ & $T_{RH}~(GeV)$ \\
     
     \hline
     0.2 & 2.955 & $7.054\times10^{13} $ \\
     
     0.4 & 2.635 & $1.693\times10^{14}$ \\
     
     0.6 & 2.426 & $2.696\times10^{14}$ \\
     
     0.8 & 2.274 & $3.669\times10^{14}$ \\
     
    \hline\hline
    
    $y=0.001$ &&\\
    \hline
     $\kappa$ & $N_{RH}$ & $T_{RH}~(GeV)$ \\
     
     \hline
     0.2 & 17.596 & $4.83\times10^{6} $ \\
     
     0.4 & 17.744 & $7.011\times10^{6}$ \\
     
     0.6 & 17.649 & $9.93\times10^{6}$ \\
     
     0.8 & 17.569 & $1.257\times10^{7}$ \\
     
    \hline
    
    \end{tabular}
    \end{adjustbox}
    \end{subtable}
\hfill
        \begin{subtable}[b]{0.48\textwidth}
\centering
     \addtolength{\tabcolsep}{-0.5pt}
        \small
        \begin{adjustbox}{max width=\textwidth}
    \begin{tabular}{|c c c|}
    \hline
    $y=10^{-6}$ &&\\
    \hline
     $\kappa$ & $N_{RH}$ & $T_{RH}~(GeV)$ \\
     
     \hline
     0.2 & 23.762 & $4.67\times10^{3} $ \\
     
     0.4 & 23.911 & $6.7\times10^{3}$ \\
     
     0.6 & 23.835 & $9.3\times10^{3}$ \\
     
     0.8 & 23.756 & $1.77\times10^{4}$ \\
     
    \hline\hline
    
    $y=10^{-12}$ &&\\
    \hline
     $\kappa$ & $N_{RH}$ & $T_{RH}~(GeV)$ \\
     
     \hline
     0.2 & 36.064 & $4.6\times10^{-3} $ \\
     
     0.4 & 36.224 & $6.5\times10^{-3}$ \\
     
     0.6 & 36.145 & $9.13\times10^{-3}$ \\
     
     0.8 & 36.056 & $1.15\times10^{-2}$ \\
     
    \hline
    
    \end{tabular}
    \end{adjustbox}
    \end{subtable}
    \caption{Values of e-folds of reheating $N_{RH}$ and reheating temperature $T_{RH}$ when $\rho_\phi=\rho_R$ with $\Gamma=\Gamma(t)$ and $n=3$. }
    \label{tab:para}
\end{table}

\subsection*{Inflation-Reheating Consistency Check}

In order to ensure the internal consistency of the inflationary and reheating sectors, we explicitly verify that the spectral index $n_s$ obtained from slow-roll inflation matches the value implied by the reheating consistency condition. 

For a given choice of the inflationary parameter $\kappa$, the spectral index $n_s$ uniquely determines the field value $\phi_k$ at horizon exit and, consequently, the inflationary e-folding number $N_k$ through the slow-roll relations. Independently, the reheating framework relates $N_k$ to the reheating duration $N_{\rm RH}$ and temperature $T_{\rm RH}$ via
\begin{equation}
N_k = \ln\!\left[\left(\frac{43}{11 g_{\rm re}}\right)^{1/3}
\frac{a_0 T_0}{k} H_k 
\exp(-N_{\rm RH}) 
\frac{1}{T_{\rm RH}}\right],
\end{equation}
where $H_k$ is the Hubble parameter at horizon exit and $g_{\rm re}$ denotes the effective relativistic degrees of freedom.

For fixed $(\kappa, N_{\rm RH}, T_{\rm RH})$, we compute $N_k$ from the reheating relation and then determine the corresponding $n_s$ from the inflationary dynamics. We find that agreement between the two sectors occurs only for specific combinations of $(\kappa, N_k, n_s)$ within numerical tolerance. In other words, the viable parameter space is restricted to the intersection of the inflationary solution manifold and the reheating consistency surface.

\begin{figure}
    \centering
    \includegraphics[width=0.8\linewidth]{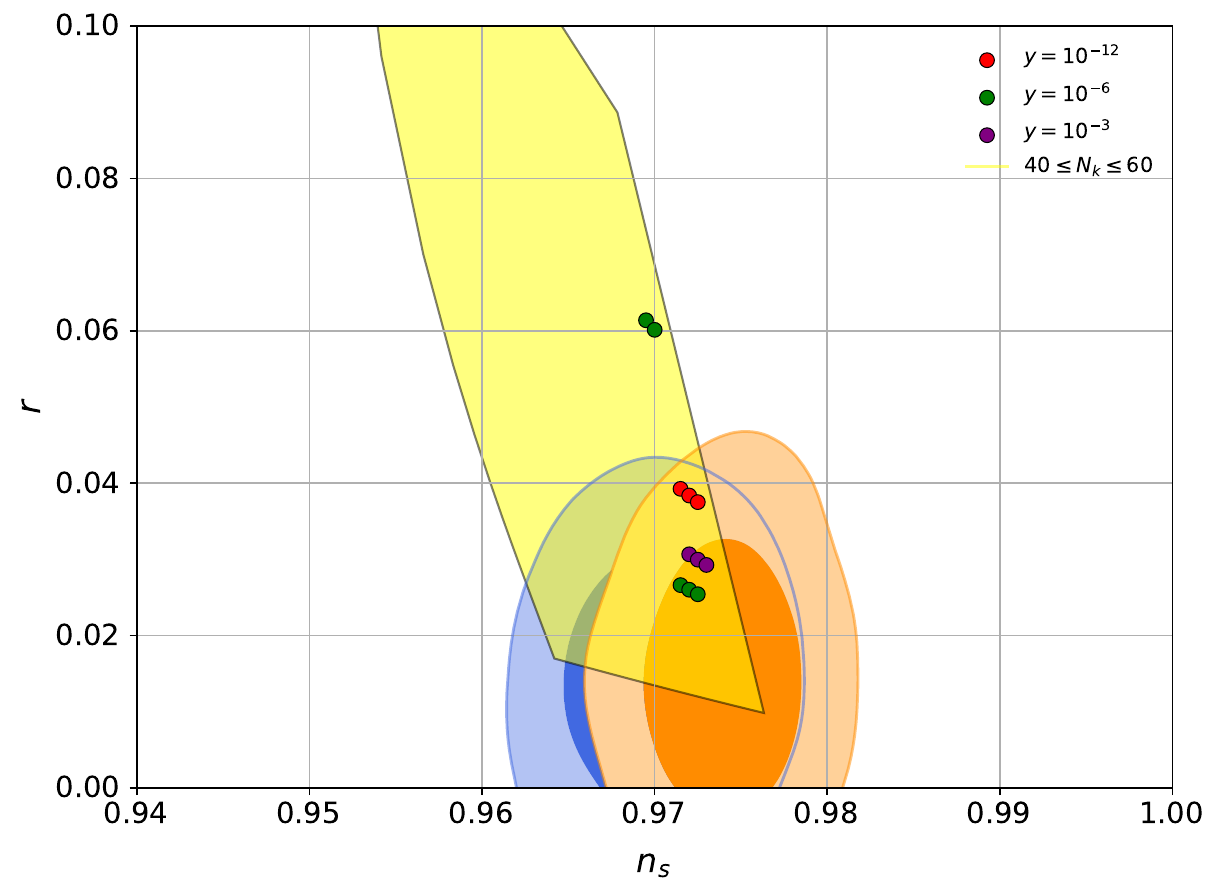}
    \caption{
    Inflation–reheating consistency in the $(n_s,r)$ plane. 
    The dark blue and orange shaded regions denote the marginalized posterior contours from the \textit{\emph{Planck}} and \textit{\emph{ACT}} $(n_s,r)$ chains, respectively. 
    The yellow band represents the continuous slow-roll inflationary predictions obtained by varying $\kappa$ and the corresponding horizon-exit e-folding number $N_k$. 
    The superimposed discrete points correspond to solutions that simultaneously satisfy the reheating consistency condition for fixed values of the Yukawa coupling $y$. 
    The restriction of the discrete reheating solutions to narrow segments of the inflationary band illustrates the reduction of the allowed parameter space once inflation–reheating matching is imposed.
    }
    \label{fig:infl_reh_consistency}
\end{figure}

The restriction of the viable parameter space is illustrated explicitly in Fig.~\ref{fig:infl_reh_consistency}. 
The yellow band represents the continuous set of slow-roll inflationary solutions in the $(n_s,r)$ plane obtained by varying $\kappa$ and the corresponding horizon-exit e-folding number $N_k$, where the left-hand boundary represents $N_k=40$ and the right-hand side boundary represents $N_k=60$. This band therefore constitutes the full inflationary solution manifold consistent with background slow-roll dynamics alone. For reference, we also display the marginalized $(n_s,r)$ posterior contours from \emph{Planck} and \emph{ACT}, which provide the observational constraints against which the inflation–reheating consistent solutions are compared. The superimposed discrete points correspond to parameter combinations that additionally satisfy the reheating consistency condition for fixed values of the Yukawa coupling $y$, which determines the decay rate and hence $(N_{\rm RH}, T_{\rm RH})$. The consistency condition is imposed by requiring
\begin{equation}
N_k^{\rm inf}(\kappa,n_s)
=
N_k^{\rm reh}(\kappa,y),
\end{equation}
within numerical tolerance. 

Operationally, for fixed $(\kappa,y)$ we scan over $n_s$, compute $N_k^{\rm inf}$ from the slow-roll solution, compute $N_k^{\rm reh}$ from the reheating evolution, and identify the value(s) of $n_s$ for which the equality holds. Only those parameter combinations satisfying this condition are retained. It is evident that the reheating solutions do not populate the entire inflationary band. Instead, they lie only at isolated locations along it. The Table  values for which the consistency holds are 
This demonstrates that enforcing the reheating consistency condition selects a lower-dimensional subset of the otherwise continuous inflationary solution space. 

\begin{table}[H]
    \centering
    
\begin{adjustbox}{max width=\textwidth}
     \addtolength{\tabcolsep}{-0.5pt}
        \small
    \begin{tabular}{|c c c c c|}
    \hline

     $y$&$\kappa$ & $N_{k}$ & $n_s$ & $r$ \\
     
     \hline
     $10^{-12}$ &0.4& 58 & $0.9720$&$0.038 $ \\
     
     $10^{-6}$ &0.2& 60 & $0.9700$&$0.060 $ \\
     
     $10^{-6}$ &0.8& 54 & $0.9720$&$0.026 $ \\
     
     $10^{-3}$ &0.6& 56 & $0.9725$&$0.030 $ \\
     
    \hline
\end{tabular}
\end{adjustbox}

    \caption{Parameter space in which inflation–reheating consistency is obeyed in the $(n_s,r)$ plane. }
    \label{cons-check}
\end{table}
We see that for Yukawa coupling $y=10^{-12}$ to $y=10^{-3}$, the reheating consistent spectral index is in the range $0.9720-0.9725$ and the tensor-to-scalar ratio is in the range $0.026-0.060$. For $y=1$ there are no points that populate the consistency parameter space.

\section{Conclusion}\label{Conclusion}
We performed a Bayesian parameter inference analysis of the inverse-tangent inflationary potential and used it to constrain the potential parameter $\kappa$ and the horizon-exit e-folding number $N_k$, together with the reheating temperature $T_{\rm RH}$ and reheating duration $N_{\rm RH}$. For a range of reheating equation-of-state parameters $\omega_{RH}$, we find that both the \textit{\emph{Planck}} and \textit{\emph{ACT}} datasets prefer $\kappa \sim 0.5$–$0.6$, although with moderate uncertainties. The corresponding values of $N_k$ favored by the data lie within the standard slow-roll expectation of $N_k = 40$–$60$, demonstrating consistency between the Bayesian constraints and the theoretical requirements of successful inflation.

In the reheating sector, we observe that for most reheating histories characterized by different values of $\omega_{RH}$, the preferred reheating temperature lies in the range $T_{\rm RH} \sim 10^{10}$–$10^{14}$ GeV. An exception arises for $\omega_{RH}=0$, where the \textit{\emph{Planck}} dataset favors a significantly lower reheating temperature, $T_{\rm RH} \approx 10^{5}$ GeV. The reheating duration correspondingly spans a broad range, $N_{\rm RH} \sim 3$–$36$ e-folds, depending on the assumed reheating equation of state.

Using these reheating-constrained posteriors, we further constructed reheating-weighted $H_0$ distributions via Monte Carlo sampling and compared them with the marginalized \textit{\emph{Planck}} and \textit{\emph{ACT}} $H_0$ posteriors. We find that the \textit{\emph{ACT}} dataset exhibits very good agreement with the reheating dynamics predicted by our model in its estimation of $H_0$. These results indicate that incorporating information about reheating histories can play an important role in the determination of the Hubble parameter and may provide additional insight into the ongoing discussion of $H_0$ measurements. Further, we also investigated reheating dynamics of inverse-tangent potential in light of the dynamical equation of state parameter $\omega_{RH}$. We studied two cases of decay rate: constant and time-dependent. The DEOS reheating framework reveals that the post-inflationary evolution is governed primarily by the microphysical decay process rather than by the detailed shape of the inflationary potential. For a constant decay rate, reheating proceeds through a smooth and continuous energy transfer from the inflaton condensate to radiation, with the effective equation of state asymptotically approaching $\omega_{RH}=1/3$ as radiation domination is established. Increasing the parameter $n$ accelerates the dilution of the inflaton energy density, leading to a shorter reheating duration and a higher reheating temperature. 

When a time-dependent decay rate $\Gamma(t)$ is considered, the reheating dynamics become explicitly sensitive to the Yukawa coupling $y$, while remaining largely insensitive to the inflationary parameter $\kappa$. Across the range $\kappa = 0.2$--$0.8$, the variation in $N_{\rm RH}$ is negligible for fixed $y$, confirming that $\kappa$ primarily controls the slow-roll phase and not the subsequent reheating epoch. In contrast, varying $y$ over several orders of magnitude produces dramatic changes in both the reheating duration and temperature. The numerical results reproduce the expected perturbative scaling $T_{\rm RH} \propto \sqrt{\Gamma M_{\rm Pl}} \propto y$, demonstrating consistency between analytic expectations and the full dynamical evolution.

The inflation–reheating consistency condition further sharpens these conclusions. While slow-roll inflation alone yields a continuous band of solutions in the $(n_s,r)$ plane corresponding to $40 \leq N_k \leq 60$, imposing reheating matching restricts the viable parameter space to isolated loci along this band. Only specific combinations of $(\kappa, y)$ yield simultaneous agreement between the inflationary prediction for $N_k$ and the reheating relation. Thus, reheating does not merely shift the preferred inflationary parameters; it reduces the dimensionality of the allowed solution manifold.

Taken together, these results demonstrate that within the DEOS framework reheating is dynamically emergent rather than externally imposed. The duration and temperature of reheating arise from the interplay between oscillatory inflaton dynamics and microphysical decay, and the requirement of inflation–reheating consistency significantly constrains the inflationary parameter space beyond slow-roll considerations alone.

\section{Acknowledgment}
MA thanks the BITS Pilani K K Birla Goa campus for the fellowship support. The work of PKD is supported by ANRF Grant No. CRG/2023/008877.

\end{document}